\documentclass[aps, prd, twocolumn, superscriptaddress, nofootinbib]{revtex4}
\usepackage{graphicx}
\usepackage{dcolumn}
\usepackage{amssymb}
\usepackage{amsmath}%
\usepackage{color}
\usepackage[normalem]{ulem}

 \newenvironment{itemz}
 {\begin{list}{$\bullet$}{\setlength{\itemsep}{0pt}}}
 {\end{list}}

\newcommand{\bitz}{\begin{itemz}}
\newcommand{\eitz}{\end{itemz}}

\def\al{\alpha}  
\def\be{\beta} 
\def\ga{\gamma}
\def\de{\delta}
\def\ep{\epsilon}

\def\la{\lambda}
\def\rh{\rho}
\def\si{\sigma}
\def\ta{\tau}

\def\De{\Delta}

\def\bi{{\mathbf{i}}}
\def\bj{{\mathbf{j}}}

\def\bv{{\mathbf{v}}}
\def\bx{{\mathbf{x}}}

\def\bB{{\mathbf{B}}}
\def\bD{{\mathbf{D}}}
\def\bE{{\mathbf{E}}}

\def\bX{{\mathbf{X}}}

\def\mcH{{\mathcal H}}
\def\mcL{{\mathcal L}}

\newcommand{\ben}{\begin{equation}}
\newcommand{\een}{\end{equation}}
\newcommand{\bea}{\begin{eqnarray}}
\newcommand{\eea}{\end{eqnarray}}
\newcommand{\ba}{\begin{array}}
\newcommand{\ea}{\end{array}}
\newcommand{\bit}{\begin{itemize}}
\newcommand{\eit}{\end{itemize}}

\def\math{\mathsurround 0pt}
\def\oversim#1#2{\lower.5pt\vbox{\baselineskip0pt \lineskip-.5pt
        \ialign{$\math#1\hfil##\hfil$\crcr#2\crcr{\scriptstyle\sim}\crcr}}}

\def\pa{\partial}

\def\half{\frac{1}{2}}

\newcommand{\vev}[1]{\left\langle#1\right\rangle}

\def\bXp{\mathaccent 19 {\bX}}

\newcommand{\Df}{\Delta f}
\def\dprime{\mathaccent"707D}

\newcommand{\LatSpa}{\De x}

\newcommand{\Ltilde}{\tilde\mcL}

\newcommand{\Lbox}{L_\text{b}}

\newcommand{\Slen}{\ell} 
\newcommand{\SlenB}{\ell_\text{B}} 
\newcommand{\SlenPhi}{\ell_\text{s}} 

\newcommand{\Swid}{r_\text{s}} 
\newcommand{\SwidS}{r_\text{s}} 
\newcommand{\SwidG}{r_\text{g}} 
\newcommand{\Sem}{T^\text{s}} 
\newcommand{\Seos}{w^\text{s}} 
\newcommand{\Srep}{\tilde{\ep}} 

\newcommand{\tCG}{\tau_\mathrm{cg}}

\newcommand{\tEnd}{\tau_\text{end}}

\newcommand{\vAv}{\bar v}
\newcommand{\vrms}{\vAv}
\newcommand{\vol}{\mathcal{V}}

\def\Xp{\mathaccent 19 X}

\begin{document}


\title{Scaling from gauge and scalar radiation in Abelian Higgs string networks}

\newcommand{\addressSussex}{Department of Physics \& Astronomy, University of Sussex, Brighton, BN1 9QH, United Kingdom}
\newcommand{\HIPetc}{\affiliation{
		Department of Physics and Helsinki Institute of Physics,
		PL 64, 
		FI-00014 University of Helsinki,
		Finland
	}}

\author{Mark Hindmarsh} 
\email{m.b.hindmarsh@sussex.ac.uk}
\affiliation{\addressSussex}
\HIPetc

\author{Joanes Lizarraga}
\email{joanes.lizarraga@ehu.eus}
\affiliation{Department of Theoretical Physics, University of the Basque Country UPV/EHU, 48080 Bilbao, Spain}
\affiliation{Department of Applied Mathematics, University of the Basque Country UPV/EHU, 48013 Bilbao, Spain}

\author{Jon Urrestilla}
\email{jon.urrestilla@ehu.eus}
\affiliation{Department of Theoretical Physics, University of the Basque Country UPV/EHU, 48080 Bilbao, Spain}

\author{David Daverio}
\email{dd415@cam.ac.uk}
\affiliation{Centre for Theoretical Cosmology, Department of Applied Mathematics and Theoretical Physics, Wilberforce Road, Cambridge CB3 0WA, United Kingdom}
\affiliation{African Institute for Mathematical Sciences, 6 Melrose Rd, Muizenberg, 7945, Cape Town, South Africa}
\affiliation{D\'epartement de Physique Th\'eorique and Center for Astroparticle Physics, Universit\'e de Gen\`eve, 24 quai Ansermet, CH--1211 Gen\`eve 4, Switzerland}

\author{Martin Kunz}
\email{martin.kunz@unige.ch}
\affiliation{D\'epartement de Physique Th\'eorique and Center for Astroparticle Physics, Universit\'e de Gen\`eve, 24 quai Ansermet, CH--1211 Gen\`eve 4, Switzerland}

\date{\today}

\begin{abstract}
We investigate cosmic string networks in the Abelian Higgs model using data from a campaign of large-scale numerical simulations on lattices of up to $4096^3$ grid points. We observe scaling or self-similarity of the networks over a wide range of scales, and estimate the asymptotic values of the mean string separation in horizon length units $\dot{\xi}$ and of the mean square string velocity $\vAv^2$ in the continuum and large time limits. 
The scaling occurs because the strings lose energy into classical radiation of the scalar and gauge fields of the Abelian Higgs model. 
We quantify the energy loss with a dimensionless radiative efficiency parameter, 
and show that it does not vary significantly with lattice spacing or string separation.
This implies that the radiative energy loss underlying the scaling behaviour is not a lattice artefact, 
and justifies the extrapolation of measured network properties to large times for computations of cosmological perturbations.
We also show that the core growth method, which increases the defect core width with time to extend 
the dynamic range of simulations, does not introduce significant systematic error.
We compare  $\dot{\xi}$ and $\vAv^2$ to values measured in simulations using the Nambu-Goto approximation, 
finding that the latter underestimate the mean string separation by about 25\%, and overestimate 
$\vAv^2$ by about 10\%.
The scaling of the string separation implies that string loops decay by the emission of massive radiation 
within a Hubble time in field theory simulations, in contrast to the Nambu-Goto scenario 
which neglects this energy loss mechanism. 
String loops surviving for only one Hubble time emit much less gravitational radiation 
than in the Nambu-Goto scenario, 
and are consequently subject to much weaker gravitational wave constraints on their tension. 
\end{abstract}


\maketitle


\section{Introduction}
\label{sec:intro}

Field theory simulations have successfully been used for many years to calculate the power spectrum of Cosmic Microwave Background (CMB) fluctuations from cosmic strings \cite{Bevis:2006mj,Bevis:2010gj,Lizarraga:2016onn}.  Searching for the imprint of strings is a powerful way to search for phase transitions just below the energy scale of inflation, which could be as high as the Grand Unification scale. As yet, no such imprints have been found \cite{Ade:2015xua}, which puts important constraints on the history of the universe and the sequence of symmetry-breaking.

A crucial feature of cosmic string networks is scaling 
\cite{Kibble:1976sj,Zeldovich:1980gh,Vilenkin:1981kz,VilShe94,Hindmarsh:1994re,Copeland:2011dx,Hindmarsh:2011qj}  
which stipulates how the energy-momentum tensor depends on the symmetry-breaking scale $\phi_0$ (inversely proportional to the string width $\Swid$), 
and certain global properties of the network: the mean string separation $\xi$ and the mean square velocity $\vAv^2$. 
In a scaling network, the string separation increases in proportion to the horizon length, 
so that $\dot\xi$ is a constant, 
while the mean square velocity $\vAv^2$ is constant. 
The mean square  fluctuations in the energy density, pressure, shear and isotropic stresses go as $\phi_0^4/\xi^4$, 
while the mean square momentum density goes as $\vAv^2\phi_0^4/\xi^4$. 
 
The components of the energy-momentum tensor are sources for gravitational perturbations, which in turn produce characteristic fluctuations in the CMB.  Calculating the most accurately measured observables, the CMB temperature and polarisation power spectra, requires the unequal time correlators (UETCs) of the energy momentum tensor, which currently can only be computed numerically although many of their properties can be understood on the basis of relatively simple modelling \cite{Vincent:1996qr,Albrecht:1997mz,Pogosian:1999np,Avgoustidis:2012gb}.
In this context, scaling is crucial: it means that numerical simulations at accessible values of $\Swid/\xi$ can be extrapolated to the much smaller values relevant for calculations of the CMB perturbations.

It is therefore clearly important to establish that the UETCs scale, a difficult computational task. 
Evidence for scaling in the mean string separation in Abelian Higgs model has been around for some time \cite{Vincent:1997cx}.
The equal time correlators (ETCs) have also been shown to scale \cite{Bevis:2010gj,Daverio:2015nva}.
However, it is also important to be sure that 
$\dot\xi$ 
and the value of $\vAv$ are reliably measured, as the overall normalisation of the Fourier-space UETCs (and therefore the CMB power spectrum) 
is proportional to $\xi^{-1}$, and in the case of the vector UETC, $\vAv^2$ also. 
The first purpose of this paper is to investigate the accuracy of the most recent simulations \cite{Daverio:2015nva} in this regard, and to estimate from a finite size analysis the asymptotic scaling values of $\dot\xi$ and $\vAv$. 
Our results for the radiation and matter eras are shown in the first two data columns of Table \ref{t:final_fits}.

The maintenance of scaling requires that the energy in the string network is converted into radiation of some kind, at a rate which also scales, which means that the energy loss rate per unit volume goes as $\mu/\xi^3$.  The constant of proportionality, which we call the string radiative efficiency, is another fundamental parameter of a string network, although not independent as it is related to $\dot\xi$ and $\vAv$ by conservation of energy. 
We measure the radiative efficiency of Abelian Higgs strings for the first time, demonstrating that it is approximately constant over almost an order of magnitude in $\Swid/\xi$, and providing asymptotic values in the last column in Table \ref{t:final_fits}.

In order to do this, it is important that $\xi$ is determined from the rest frame string length, which is proportional to the total energy in string. 
Estimating the rest-frame string length from a field theory simulation is not trivial, and has not been done before, except for field configurations without radiation \cite{Moore:2001px}.  
We introduce new ways of measuring the rest-frame string length in Section \ref{s:AHprops}, testing them against standing wave configurations where the rest-frame length is very clearly conserved.  The same field configurations allow us to test, correct and extend estimators for the mean square string velocity developed in Ref.~\cite{Hindmarsh:2008dw}. 

For our largest simulations, $\Swid/\xi \simeq 0.003$, 
and so a widely-given argument would say that the strings should obey the Nambu-Goto equations of motion \cite{Forster:1974ga,Anderson:1997ip}, as the local string curvature should also be of order $\xi$.
In the Nambu-Goto description, conservation of energy is equivalent to conservation of rest-frame string length, with corrections which decrease exponentially with $\xi/\Swid$, due to perturbative classical radiation of massive gauge and Higgs modes \cite{Olum:1999sg}.
The fact that string length is not conserved even for $\Swid/\xi \ll 1$  means that there is another radiation mechanism, 
not captured in the Nambu-Goto approximation, which causes string loops to decay within a Hubble time.
Unless this mechanism has some very weak dependence on $\Swid/\xi $ unresolved in our simulations, 
and so eventually shuts off, we can conclude that 
simulations using the Nambu-Goto approximation \cite{Martins:2005es,Ringeval:2005kr,BlancoPillado:2011dq,Blanco-Pillado:2013qja,Blanco-Pillado:2015ana} greatly overestimate the density of string loops in the universe, and 
thus the gravitational radiation the loops produce.  

The independence of the radiative efficiency from $\xi$ is rather remarkable, 
and it is not clear how radiation of frequency $\Swid^{-1}$ is generated by a 
system whose dynamical timescale is $\xi$.
Some preliminary ideas on the mechanism were given in Ref.~\cite{Hindmarsh:2008dw}, 
where it was pointed out that the presence of small-scale structure on the strings automatically puts high-frequency power into the field. 
Simulations of domain walls in two and three dimensions \cite{Press:1989yh,Garagounis:2002kt,Borsanyi:2007wm,Leite:2011sc,Martins:2016ois} 
also point to a similar mechanism at work.
In this paper we restrict ourselves to measuring the efficiency parameter and demonstrating that 
it is approximately scale-independent, and examining the lattice effects in much greater detail than before. 
We also show that modifying the field equations to allow string cores to grow with time 
\cite{Press:1989yh,Bevis:2006mj}
does not significantly affect the radiation mechanism.
A more detailed investigation of small-scale structure, and of the lifetime of string loops in our simulations, will be reported on elsewhere.


\section{Abelian Higgs model in a FLRW background}

The simplest field theory with string defects is the Abelian Higgs model, which has been 
extensively used to study the dynamics of cosmic strings. The next simplest is an SU(2) 
gauge theory with two adjoint scalars, which has recently been studied numerically for the first time \cite{Hindmarsh:2016lhy,Hindmarsh:2016dha}.

In this section we recall the Abelian Higgs model in the classical approximation, 
which gives the leading terms in the energy-momentum correlations, at least 
at weak coupling. We give the continuum and lattice formulations in an expanding 
universe, and describe the core growth method used to extend the dynamic range.

\subsection{Continuum formulation}
The action for the Abelian Higgs model in a general spacetime background is
\bea
S &=& -\int d^4x \, \sqrt{-g} \Big( g^{\mu\nu}D_\mu\phi^*D_\nu\phi +V(\phi) \nonumber\\
&& + \frac{1}{4e^2}g^{\mu\rho}g^{\nu\si}F_{\mu\nu}F_{\rho\si}\Big),
\eea
where $\phi(x)$ is a complex {scalar} field,  
$A_\mu(x)$ is a {vector} field, 
the covariant derivative is \(D_\mu = \partial_\mu - iA_\mu\), 
and the potential is \(V(\phi) = \half\la(|\phi|^2 - \phi_0^2)^2\).
For the Friedmann-Lema\^itre cosmology in the early universe we use the spatially flat Robertson-Walker (FLRW) metric  
\ben
g_{\mu\nu} = a^2(\ta)\eta_{\mu\nu}
\een
where $\eta_{\mu\nu} = \text{diag}(-1,1,1,1)_{\mu\nu}$ is the Minkowski metric, 
$\ta$ is conformal time, and $a(\ta)$ is the scale factor, 
proportional to \(\ta\) and \(\ta^{2}\) in the radiation and matter eras respectively.

The resulting field equations in the temporal gauge ($A_0 = 0$) are 
\bea
\ddot\phi + 2\frac{\dot a }{a}\dot \phi -\bD^2\phi + a^2\lambda (|\phi|^2 -v^2)\phi &=& 0, \\
\pa^\mu\left(\frac{1}{e^2} F_{\mu\nu} \right) - ia^2(\phi^*D_\nu\phi - D_\nu\phi^*\phi) &=& 0,
\eea
where indices are raised with the Minkowski metric.
Variation with respect to $A_0$ gives Gauss's Law, 
\ben
\pa_i  E_i =  i e^2 a^{2} (\dot\phi^*  \phi - \phi^* \dot \phi  ),
\label{e:GauLaw}
\een
where $E_i = F_{0i}$.

It is most convenient to discretise the comoving coordinates, but this presents a problem.  
\ben
\label{e:SwidDef}
\SwidS = (\sqrt{\la/2}\phi_0)^{-1}, \quad
\SwidG = (e\phi_0)^{-1},
\een
shrink in comoving coordinates. 
As these length scales control the width of the string, they must be kept greater than the lattice spacing, otherwise lattice artefacts can become serious. 

A solution is to modify the equations so that the coming length scales do not shrink so fast 
\cite{Press:1989yh,Bevis:2006mj} or equivalently that the physical width of string core grows, to read 
\bea
\ddot\phi + 2\frac{\dot a }{a}\dot \phi -\bD^2\phi + \lambda_0 a^{2s}(|\phi|^2 -\phi_0^2)\phi &=& 0,\nonumber \\
\pa^\mu\left(\frac{a^{2(1-s)}}{e_0^2} F_{\mu\nu} \right) - ia^2(\phi^*D_\nu\phi - D_\nu\phi^*\phi) &=& 0.\nonumber
\eea
One can view this change as the introduction of time-dependent couplings 
 \ben
  e^2(\ta) = e^2_0a^{2(s-1)}(\ta), \quad \la(\ta) = \la_0a^{2(s-1)}(\ta),
\een
keeping the vacuum expectation value $\phi_o$ fixed.
The time-dependence of the couplings is the same, so that the string tension
\ben
\label{e:StrTen}
\mu = 2\pi\phi_0^2 B(\beta)
\een
is unchanged, where $B(\beta)$ is a slowly-changing function of $\be = \sqrt{2\la}/e$, and $B(1) = 1$ \cite{Hill:1987qx}. The choice $0\le s < 1$ has the effect of increasing the physical string width with time.
When $s=0$, the string width is fixed in comoving coordinates. The true dynamics are recovered at $s=1$.
Gauss's Law is preserved for any value of $s$, but energy-momentum conservation is generally spoiled for $s < 1$. In practice, the violations are of order a few per cent.

Throughout this work, and in previous ones (for example \cite{Bevis:2006mj,Hindmarsh:2008dw,Bevis:2010gj,Daverio:2015nva,Lizarraga:2016onn}), the critical coupling $\beta=1$ has been chosen. At the critical coupling, parallel straight strings in Minkowski space do not interact, whereas for $\beta<1$ strings attract each other, and for $\beta>1$ they repel. In the $\beta<1$ case strings can form bound states due to their attraction, and the 
network scaling properties 
could be somewhat different from the $\beta=1$ case, as pointed out in \cite{Hiramatsu:2013tga}.  We will report on results of our simulations for $\beta<1$ in a future publication.


\subsection{Lattice formulation}

We begin with the continuum Hamiltonian density in the temporal gauge
\bea
\mcH &=& \frac{1}{2e^2(\ta)} \left(\bE^2  +  \bB^2 \right) + a^2\left(|\pi|^2 + |\bD\phi|^2\right)  \nonumber\\
& &  + \frac14 a^4 \la(\ta)(|\phi|^2 - \phi_0^2)^2,
\label{e:HamDen}
\eea
where $E_i = F_{0i}$ and $B_i = \half\ep_{ijk}F_{jk}$. 

We discretise the field equations in the standard way \cite{Moriarty:1988fx,Bevis:2006mj}, defining complex variables $\phi_{\bx}(\ta)$ on the sites of a regular cubic periodic lattice with spacing $\LatSpa$, and real variables $\theta_{i,\bx}(\ta)$ on the links in the positive $i = 1,2,3$ directions at site $\bx$.
With canonical momenta  \(\pi_{\bx}(\ta) \) and \(\ep_{i,\bx}(\ta)\), the terms in the Hamiltonian are represented as follows:
\bea
|\bD\phi(x)|^2  & \to & 
\frac{1}{{\De x}^2}\sum_i|e^{-i\theta_{i,\bx}}\phi_{\bx+\bi} - \phi_\bx|^2,
\\
\frac{1}{2} \bB^2(x) & \to & \frac{1}{2{\De x}^4}\sum_{\langle i,j\rangle}\left[ 1 - 
\cos(\Box_{ij,\bx}) 
\right], \label{flux}
\\
|\pi(x)|^2  & \to &  | \pi_\bx |^2 ,  
\\
\frac{1}{2} \bE^2(x) & \to & \frac{1}{2{\De x}^4}\sum_{i} \ep_{i,\bx}^2,
\eea
where $\Box_{ij,\bx} = \theta_{i,\bx}+\theta_{j,\bx+\bi}-\theta_{i,\bx+\bj}-\theta_{j,\bx}$ is the flux through the plaquette $\vev{i,j}$ at $\bx$. The lattice representation of the potential energy density $V$ is obvious.

Time evolution is performed through a leapfrog
\begin{eqnarray}
\phi_\bx^n &=& \phi_\bx^{n-1} + \pi_\bx^{n-\half} \De t, \\
(a^2\pi_\bx)^{n+\half} &=& (a^2\pi_\bx)^{n-\half} + F_\bx^n  \De t, \\  
\theta_{i,\bx}^n &=& \theta_{i,\bx}^{n-1} + \ep_{i,\bx}^{n-\half} \De t, \\
\left(\frac{\ep_{i,\bx}}{e^2}\right)^{n+\half} &=& \left(\frac{\ep_{i,\bx}}{e^2}\right)^{n-\half}  + G_{i,\bx}^n \De t ,
\eea

where the force terms are 
\bea
F_\bx^n &=& \frac{\pa {\cal H}_\bx }{ \pa {\phi}^n_{\bf x}}, \\
G_{i,\bx}^n &=&  \frac{\pa {\cal H}_\bx }{ \pa {\theta}^n_{i,\bf x}}. 
\eea

The scheme is O$(\De x^2,\De t^2)$ accurate, and when $a$ is constant, conserves a quantity which converges to the total energy in the continuum limit.
It is possible to improve the spatial derivative terms to O$(\De x^4)$ accuracy straightforwardly 
\cite{Hindmarsh:2014rka} but improvement of the gauge field time derivative requires an implicit update \cite{Moore:1996wn}.


\subsection{Energy, Lagrangian and pressure}

The symmetrised energy-momentum tensor for the Abelian Higgs model is
\bea
T_{\mu\nu} &=& - \frac{1}{e^2} F_{\mu\rho}F_{\nu\si}g^{\rho\si} \nonumber\\
&-& D_\mu\phi^*D_\nu\phi - D_\nu\phi^*D_\mu\phi + g_{\mu\nu}\mcL.
\eea
Note that $T_{00} = - a^{-2}\mcH$, and that
the physical energy density is $\rho = {T^0}_{0} = a^{-4}\mcH$. 
It will be useful to label the individual terms in the energy density as in Table \ref{t:LagMon}.

\begin{table}
\begin{tabular}{ccccc}
\hline
\hline
\(\rho_E\) & 
\(\rho_B\) & 
\(\rho_\pi\) & 
\(\rho_D\) & 
\(\rho_V\) \\
\hline\\[-6pt]
\( \displaystyle {\frac{\bE^2}{2e^2a^4}} \) &
\(\displaystyle{\frac{\bB^2}{2e^2a^4}} \) &
\(\displaystyle \frac{|\pi|^2}{a^2} \) &
\(\displaystyle \frac{|\bD\phi|^2}{a^2} \) &
\(\displaystyle{ V(\phi)} \) \\[6pt]
\hline
\hline
\end{tabular}

\caption{\label{t:LagMon}
Definitions of symbols used to denote terms in the energy density, pressure and Lagrangian density.
}
\end{table}

With these definitions, we can write the energy density, pressure ($p = - {T^i}_{i}$) and Lagrangian density as 
\bea
\rho &=& \rh_E + \rh_B + \rh_\pi + \rh_D + \rh_V,  \\
p &=& \frac{1}{3}\rh_E + \frac{1}{3}\rh_B + \rh_\pi - \frac{1}{3}\rh_D - \rh_V,  \\ 
\mcL &=& \rh_E - \rh_B + \rh_\pi - \rh_D - \rh_V.
\eea

We denote volume-integrated quantities 
\ben
E_{X} = a^2\int d^3x\, \rho_X.
\een
The extra factor of $a^2$ is for later convenience.


\section{Abelian Higgs string properties}
\label{s:AHprops}

\subsection{Nielsen-Olesen vortex}

In Minkowski space-time, which can be recovered by setting $a=1$,  
the field equations have static cylindrically symmetric solutions, Nielsen-Olesen (NO) vortices \cite{Nielsen:1973cs}. 
A suitable Ansatz is 
\ben
\phi = \phi_0 f(x)e^{i\theta}, \quad A_i = \hat\varphi_i \frac{\phi_0 g(x)}{x}, 
\label{e:NOAns}
\een
where $x = r / \SwidG$, $r$ is a radial cylindrical coordinate, and $\varphi$ the angular coordinate.

The functions $f$ and $g$ approach the vacuum outside the string as
\bea
f &\simeq &
          1-f_1 x^{-1/2}\exp(-\sqrt{\beta} x), \\
g &\simeq& 1-g_1x^{1/2}\exp(-x), 
\label{e:NOAsy}
\eea
where $f_1$ and $g_1$ are constants.
Hence $\SwidS$ and $\SwidG$ control the approach of the scalar and gauge fields to the vacuum.

In an expanding universe, these solutions are a very good approximation, as corrections are of order  
$H\SwidS$ and $H\SwidG$, where $H = \dot a /a$ is the Hubble rate in conformal time.

In the NO solution,  \(\rho\), \(p\) and \( \mcL\) are all strongly peaked at the origin, giving rise to a tube-like concentration of energy, pressure and Lagrangian density (see Fig.~\ref{f:NOslice}).  

\begin{figure}[h]

\includegraphics[width=0.5\textwidth]{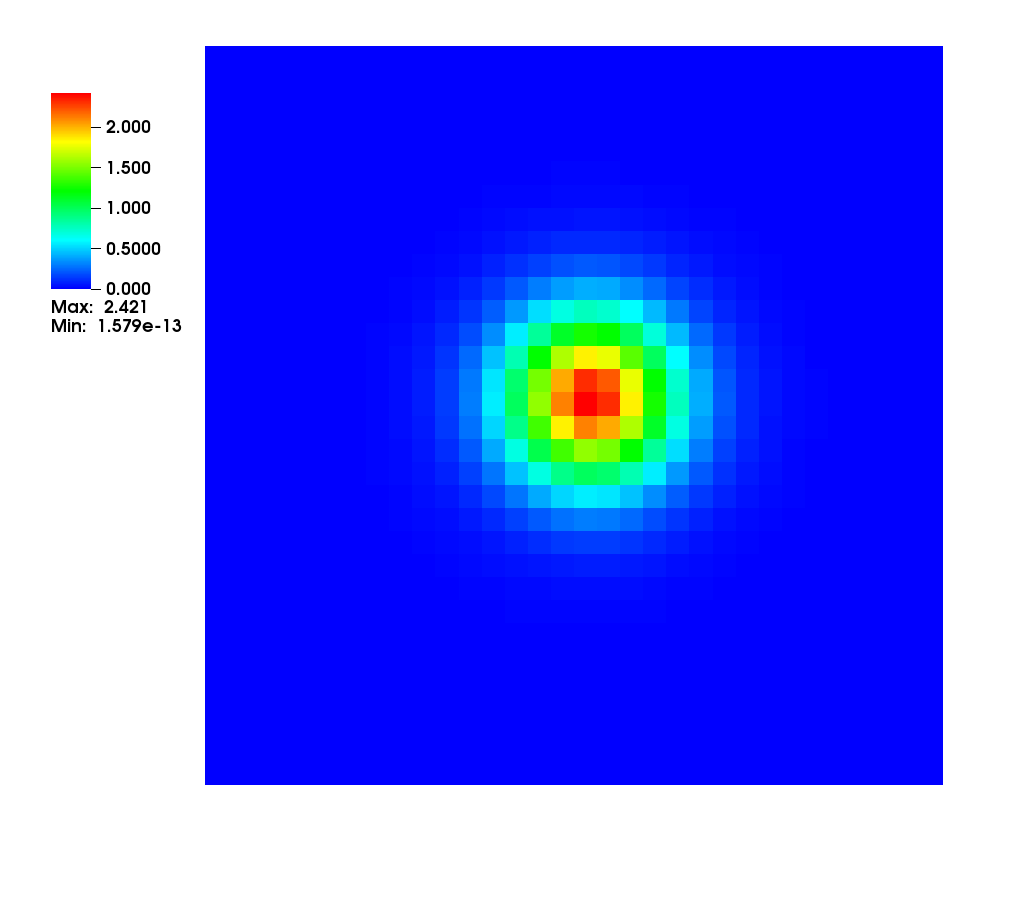}

\caption{
Slice through the Nielsen-Olesen vortex, showing total energy density $\rho$, in a $64\times64$ region with $\LatSpa = 0.25$.
The total energy density is equal to the negative of the Lagrangian density and the negative of the diagonal component orthogonal to the plane, $T_{33}$.
}

\label{f:NOslice}
\end{figure}

Writing the NO field configurations as $\phi_\text{s}$, $\bB_\text{s}$, 
noting that $\bB_\text{s}$ is orthogonal to $\bD\phi_\text{s}$, and choosing coordinates so the $\bB_\text{s}$ is in the $x_3$ direction, 
the non-zero components of the energy-momentum tensor are
\bea
T_{00} &=& -a^2\left( V(\phi_\text{s}) + \frac{1}{a^2} |\bD\phi_\text{s}|^2 + \frac{1}{2e^2a^4} \bB^2_\text{s} \right), \\
T_{33} & = & a^2\left( V(\phi_\text{s}) + \frac{1}{a^2} |\bD\phi_\text{s}|^2 + \frac{1}{2e^2a^4} \bB^2_\text{s} \right), \\
T_{ab} & = & a^2\left( V(\phi_\text{s})  -  \frac{1}{2e^2a^4} \bB^2_\text{s} \right) \de_{ab}.
\eea
Hence we see that the pressure along the string $-{T^3}_3$ is equal and opposite in sign to the energy density ${T^0}_0$.  
Note that, as a consequence of Derrick's theorem \cite{Derrick:1964ww,Manton:2004tk}, the planar components of the energy-momentum tensor vanish when integrated across the plane, $\int d^2 x T_{ab}=0$.

\subsection{Nambu-Goto strings}
\label{ss:MovNO}

In the next section we will show that a field configuration which is locally a NO vortex behaves like a Nambu-Goto (NG) string. In this section we will briefly recall some useful details about NG string dynamics. 
For simplicity we study only the motion in Minkowski space in detail.

First, let us recall the Nambu-Goto action 
\ben
S_\text{NG} = -\mu \int d^2\si \sqrt{-\ga},
\een
where $\mu$ is the string tension,
$\si^\al$ ($\al=0,1$) are string world sheet coordinates, 
and $\ga_{\al\be} = \pa_\al X^\mu \pa_\be X^\nu g_{\mu\nu}$ is the embedding metric.  
In the standard gauge, the worldsheet time $\si^0$ is identified with 4-dimensional time $X^0$, and the constraints
\bea
\dot\bX\cdot\bXp &=& 0,  \label{e:NGcon1}\\
\dot\bX^2 + \bXp^2 &=& 1, \label{e:NGcon2}
\eea
are chosen. Here, the prime denotes the derivative with respect to the spacelike worldsheet coordinate $\si^1$. 
The equation of motion is then 
\ben
\ddot\bX - \dprime{\bX} = 0. 
\een

It will be useful to write $\si$ for the spacelike worldsheet coordinate, and note that in the standard gauge the Nambu-Goto Lagrangian can be written 
\ben
L_\text{NG} = -\mu  \int d\si |\bXp| \sqrt{1 - \dot\bX^2}.
\een
Using the gauge condition (\ref{e:NGcon2}), we have 
\ben
L_\text{NG} = -\mu \int d\si \left({1 - \dot\bX^2}\right).
\een
The energy-momentum tensor of the NG string in Minkowski space is is 
\ben
T^{\mu\nu}(\bx,t) = \mu \int d\si \left( \dot X^{\mu}\dot X^\nu - {\Xp}^\mu  {\Xp}^\nu   \right) \de(\bx - \bX(\si,t)).
\label{e:NGeneMom}
\een
Hence the total energy of NG string is
\ben
E_\text{NG} = \mu \int d^3x {T^0}_0 = \mu \Slen,
\label{e:NGene}
\een
where $\ell = \int d\si$ is the length of string in its local rest frame.
Hence  
\ben
L_\text{NG} = - E_\text{NG} \left( 1 - \vAv^2\right)
\label{e:NGlag}
\een
where 
\ben
\vAv^2 =  \frac{1}{\Slen}\int d\si \dot\bX^2
\een
defines the mean square velocity. 

 From the space-space components of the energy-momentum tensor (\ref{e:NGeneMom})
 the average string pressure is seen to be
\ben
p_\text{NG} = \frac{\mu}{3\vol} \int d\si \left(2 \dot\bX^2 - 1 \right) = \frac{E_\text{NG}}{3\vol} \left(2 \vAv^2 - 1 \right),
\label{e:NGpre}
\een
where $\vol$ is the spatial volume.

Note that the rest length (sometimes called the invariant length) does not change during the evolution of the string, as a consequence of energy conservation.

\subsection{Moving Abelian Higgs strings}

The example of NG strings will give us a method of measuring the length of string and the mean square velocity from the fields of the Abelian Higgs model. 
Estimates for the rest length and the mean square velocity were first introduced in Ref.~\cite{Moore:2001px}, which however worked only for field configurations without radiation. 

A technique for estimating the mean square velocity for strings and radiation was proposed in Ref. \cite{Hindmarsh:2008dw}, exploiting the fact that in a moving string solution, the electric field $\bE$ and the canonical scalar momentum $\pi$ are given by Lorentz boosts of the static field configuration, and weighting the field estimates so that they were preferentially taken near the string.
Here we will extend the technique to include string length estimators, and correct errors in the velocity estimators in the analysis of Ref.~\cite{Hindmarsh:2008dw}.

Let us first suppose that all the energy in the field is in the form of NO vortices. 
Denoting local rest frame space coordinates $\bx_\text{s}$, and 
fields measured in the local rest frame with the subscript s,  
the fields of a piece of string moving with velocity $\dot \bX$ orthogonal to its magnetic field are  
\bea
\bE(\bx,t) &=& \ga \dot\bX \times \bB_\text{s}(\bx_\text{s}), \\
\bB(\bx,t) &=& \ga \bB_\text{s}(\bx_\text{s}), \\
\pi(\bx,t) &=& \ga \dot\bX \cdot \bD \phi_\text{s}(\bx_\text{s}), \\
\bD\phi(\bx,t) &=&  \ga \hat{\bv} (\hat{\bv} \cdot \bD \phi_\text{s}(\bx_\text{s}) ) + \bD^\perp\phi(\bx,t),
\eea
where $\hat\bv$ is a unit vector in the direction of the velocity, $\ga =1/\sqrt{1 - \dot\bX^2}$ is the boost factor,
and 
\ben
D^\perp_i\phi(\bx,t) = (\de_{ij} - \hat{v}_i\hat{v}_j) D_j \phi(\bx,t).
\een

Hence the total energy in the electric field is 
\ben
E_E = \half \int d^3 x \bE^2 = \half \int d^3x_\text{s} \ga \bB^2_\text{s}(\bx_\text{s})  \dot\bX^2,
\een
where we have used $\bx_\text{s} = \bx^\perp + \hat{\bv}\ga(\hat{\bv}\cdot\bx - v t)$, with $\bv \equiv \dot\bX$.
Without loss of generality, we can suppose that the string is locally in the $z$ direction, and therefore that $dz_\text{s} = |\bX'| d\si = \ga^{-1} d\si$. Hence
\ben
E_E = \half \int dx_\text{s}dy_\text{s} \bB^2_\text{s}(\bx_\text{s}) \int d\si  \dot\bX^2.
\een
The total energy in the magnetic field can similarly be found as 
\ben
E_B = \half \int dx_\text{s}dy_\text{s} \bB^2_\text{s}(\bx_\text{s}) \int d\si.
\een
We can also get estimators from the  kinetic and gradient energies of the scalar field
\ben
E_{\pi} =  \int d^3 x |\pi|^2, \quad
E_{D} =  \int d^3 x |\bD\phi|^2.
 \een
A boosted string oriented in the $z$ direction has scalar kinetic energy
\bea
E_{\pi} &=&   \int dx_\text{s}dy_\text{s} D_i\phi^*_\text{s}(\bx_\text{s}) D_j\phi_\text{s}(\bx_\text{s}) \int d\si \dot X^i \dot X^i \nonumber\\
&=& \half \int dx_\text{s}dy_\text{s} |\bD\phi_\text{s} |^2 \int d\si \dot\bX^2.
\eea
The gradient energy is 
\bea
E_{D} &=&   \int dx_\text{s}dy_\text{s} \int d\si \left( |\hat{\bv} \cdot \bD\phi_\text{s} |^2  + \frac{1}{2\ga^2} |\bD\phi_\text{s} |^2 \right) \nonumber\\
&=& \half \int dx_\text{s}dy_\text{s}|\bD\phi_\text{s} |^2  \int d\si \left( 1  + \frac{1}{\ga^2} \right).
\eea
Finally, we may also write the potential energy
\ben
E_V = \int dx_\text{s}dy_\text{s}V(\phi_\text{s})  \int d\si \frac{1}{\ga^2} .
\een
We can explicitly calculate the energy, the Lagrangian, and the pressure multiplied by the volume, 
\bea
E &=& E_E + E_B + E_\pi + E_{D} + E_V, \\
L &=& E_E- E_B + E_\pi - E_{D} - E_V, \\
p\vol &=& \frac{1}{3}(E_E + E_B) + E_\pi - \frac{1}{3}E_{D} - E_V,
\eea
finding
\bea
E &=& \mu \left(1 + \Df \vAv^2\right) \Slen,  \label{e:AHene} \\
L &=& -\mu ( 1 - \vAv^2)\Slen, \label{e:AHlag}\\
p\vol &=& \frac{1}{3}\mu \left[( 2 \vAv^2 -1 ) + \Df (2 - \vAv^2)\right]\Slen , \label{e:AHpre}
\eea
where
\bea
\mu &=&  \int dx_\text{s}dy_\text{s} \left( \half \bB^2_\text{s} + |\bD\phi_\text{s}|^2 + V(\phi_\text{s}) \right),
\label{e:StrMu}
\eea
and 
\bea
\mu\Df &=& \int dx_\text{s}dy_\text{s} \left( \half \bB^2_\text{s} - V(\phi_\text{s}) \right).
\eea
Note that $\mu$ is the mass per unit length of a static string (\ref{e:StrTen}). 
We will denote each of the three contributions to the mass per unit length $\mu f_B$, $\mu f_D$, and $\mu f_V$, so that $\Df = f_B - f_V$. Their values, along with $\mu/\phi_0^2$, are shown in Table \ref{t:StrEne}, computed numerically 
for an Abelian Higgs string at critical coupling ($\la = 2e^2$) and relevant lattice spacings.
Note that as $\LatSpa\to0$, $\mu \to 2\pi$ and $\Df = f_B-f_V \to 0$ (see e.g.~\cite{Manton:2004tk}).

\begin{table}
\begin{tabular}{l|lll}
$\phi_0\LatSpa$ & $0.5$ & $0.25$ & $0.125$ \\
\hline
$\mu/\phi_0^2$ & 6.205 & 6.265 & 6.278 \\
$f_B$ & 0.215 & 0.209 & 0.208 \\
$f_D$ & 0.584 & 0.584 & 0.585 \\
$f_V$ & 0.201 & 0.206 & 0.207 \\
\end{tabular}
\caption{\label{t:StrEne}
	String tension $\mu$ and fractions due to magnetic, scalar gradient and scalar potential terms in (\ref{e:StrMu}) at critical coupling ($\la = 2e^2$), for lattice spacings used in this work. 
	Note that as $\LatSpa\to0$, $\mu \to 2\pi\phi_0^2$ and $\Df = f_B-f_V \to 0$. Units are set by the scalar expectation value $\phi_0$.
	}
\end{table}

\subsection{Field estimators for length and velocity}

For simplicity, we will first neglect the numerical corrections due to $\Df$, which are of order $5\%$.  
The first rest length estimator \cite{Moore:2001px} is then simply
\ben
\ell = \frac{E}{\mu}. 
\een
We can obtain other length estimators from the components of the energy: first, the magnetic field energy gives 
\ben
\label{e:SlenBDef}
\SlenB = \frac{E_B}{\mu f_B},
\een
and the scalar field gradient energy gives 
\bea
\SlenPhi = \frac{E_{D}+ E_{\pi}}{\mu f_D}.
\eea
There is also a combined length and velocity estimator from the potential and the electric field,
\ben
\Slen_{V,E} = \frac{E_E + E_V}{\mu f_V}.
\een
Note that the length estimators are not independent, as 
\ben
\Slen = f_B\SlenB + f_{D}\SlenPhi + f_V \Slen_{V,E}.
\een

The first velocity estimator \cite{Moore:2001px} follows from the Lagrangian,
\ben
\label{e:velLag}
\vAv^2 = 1 + \frac{L}{E}.
\een
We can construct a second velocity estimator with the 
pressure and energy density of the fields, $\bar{p}$ and $\bar{\rho}$, 
\ben
\vAv^2_w = \half\left( 1 + 3 w \right),
\een
where $w = \bar{p}/\bar{\rho}$ is the equation of state parameter.
Other velocity estimators follow from ratios of energies.
The ratio of the electric and magnetic energies of the gauge field gives an estimator
\ben
\vAv^2_\text{g} = \frac{E_E}{E_B},
\een
while an analogous ratio for the scalar field gives 
\ben
\vAv^2_{\text{s}} = \frac{2R_{\text{s}} }{1+ R_{\text{s}} },
\een
where 
\ben
R_{\text{s}} = \frac{E_{\pi}}{E_{D}}.
\een

Of course, field configurations may contain radiation as well, so the energy in the field will be an overestimate of the length of string. However, free propagating wave solutions have vanishing Lagrangian, and so the string configurations show a strongly negatively peaked Lagrangian density. 

If we want a more precise estimate of the string length, we can use 
Lagrangian weighting, which has the effect of preferentially selecting regions of space occupied by string. 
The natural scale for the Lagrangian density is the scalar expectation value $\phi_0$, and we will denote a dimensionless Lagrangian density by 
\ben
\Ltilde = \mcL/\phi_0^4.
\een

For example, the Lagrangian-weighted magnetic field energy is
\ben
E_{B,\mcL} =  - \frac{1}{2} \int d^3 x \bB^2 \Ltilde.
\een
The independent Lagrangian-weighted length estimators are 
\bea
\Slen_\mcL &=&  \frac{1}{\mu_\mcL}\frac{E_\mcL - \Df L_\mcL}{1 + \Df}, \label{e:ell_lag}\\
\Slen_{B,\mcL}  &=& \frac{E_{B,\mcL}}{\mu_\mcL f_{B,\mcL}} , \label{e:ell_B_lag}\\
\Slen_{\text{s},\mcL}  &=&  \frac{E_{D,\mcL} + E_{\pi,\mcL}}{\mu_\mcL f_{D,\mcL}},  \label{e:ell_s_lag}
\eea
where  
\ben
\mu_\mcL = - \int dx_\text{s}dy_\text{s} \left( \half \bB^2_\text{s} + |\bD\phi_\text{s}|^2 + V(\phi_\text{s}) \right)\Ltilde
\een
is the Lagrangian-weighted mass per unit length.
Its value, along with 
the fractional contributions of its three terms $f_{B,\mcL}$, $f_{D,\mcL}$ and $f_{V,\mcL}$ 
are given in Table \ref{t:StrEneLag}.

\begin{table}
\begin{tabular}{l|lll}
$\phi_0\LatSpa$ & $0.5$ & $0.25$ & $0.125$ \\
\hline
$\mu_\mcL/\phi_0^2$ & 6.841 & 6.970 & 7.006 \\
$f_{B,\mcL}$ & 0.215 & 0.207 & 0.206 \\
$f_{D,\mcL}$ & 0.587 & 0.589 & 0.589 \\
$f_{V,\mcL}$ & 0.198 & 0.204 & 0.205 \\
\end{tabular}
\caption{\label{t:StrEneLag}
	Lagrangian-weighted string tension $\mu_\mcL$ and fractions due to magnetic, scalar gradient and scalar potential terms in (\ref{e:StrMu}) at critical coupling ($\la = 2e^2$), for lattice spacings used in this work. 
	Note that as $\LatSpa\to0$, $\Df_\mcL = f_{B,\mcL}-f_{V,\mcL} \to 0$ at critical coupling. 
	Units are set by the scalar expectation value $\phi_0$.
}
\end{table}

The fractional contributions change by about 1\% when Lagrangian-weighted, so we neglect the distinction between $f_B$ and $f_{B,\mcL}$.

The Lagrangian-weighted velocity estimators are
\bea
\vAv^2_{\mcL} &=& \frac{E_\mcL + L_\mcL}{E_\mcL - \Df L_\mcL},  \label{e:v2_lag}\\
\vAv^2_{\text{B},\mcL} &=& \frac{E_{E,\mcL}}{E_{B,\mcL}}, \label{e:v2_B_lag}\\
\vAv^2_{\text{s},\mcL} &=& \frac{2R_{\text{s},\mcL} }{1+ R_{\text{s},\mcL} }, \label{e:v2_s_lag}
\eea
with 
\ben
R_{\text{s},\mcL} = \frac{E_{\pi,\mcL}}{E_{D,\mcL}}.
\een
One can also estimate the mean square velocity from a combination of the pressure and the energy density,
\ben
\vAv^2_{w,\mcL} = \frac{ 1 + 3 w_\mcL - 2 \Df}{2 - \Df(1+3w_\mcL)}, 
\label{e:v2_w_lag}
\een
where $w_\mcL = \bar{p}_\mcL\vol/E_\mcL$ is the string equation of state parameter, determined from Lagrangian-weighted quantities.

Eqs.\ (\ref{e:ell_lag}), (\ref{e:ell_B_lag}) and (\ref{e:ell_s_lag}) 
comprise our final length estimators, 
with Eqs.\ (\ref{e:v2_lag}), (\ref{e:v2_B_lag}), (\ref{e:v2_s_lag}) and (\ref{e:v2_w_lag}) our final mean square velocity estimators.

\subsection{Velocity estimates from string position}
\label{velfrompos}
Another estimator of the string velocities can be obtained directly from the positions of the strings. This estimator demands a higher level of modelling since, contrary to previous estimators,  it is not local in either space or time.
Instead, it relies on the comparison of positions of the string cores at different times, with the positions determined from the gauge-invariant ``winding" of the phase of the scalar field around a plaquette on the lattice. 
A similar method has been used for estimating monopole velocities \cite{Lopez-Eiguren:2016jsy}. 

This procedure has its own advantages and disadvantages. 
On the one hand, we deal only with string positions, and therefore we calculate the string velocities without any pollution from the radiation that is present in the simulation. 
This procedure   also seems easier to compare to Nambu-Goto simulations, as we have direct information about the string positions. On the other hand, as mentioned before, the procedure for getting the velocities of strings is more complicated, increasing the uncertainties.

\begin{figure}
\includegraphics[width=0.5\linewidth]{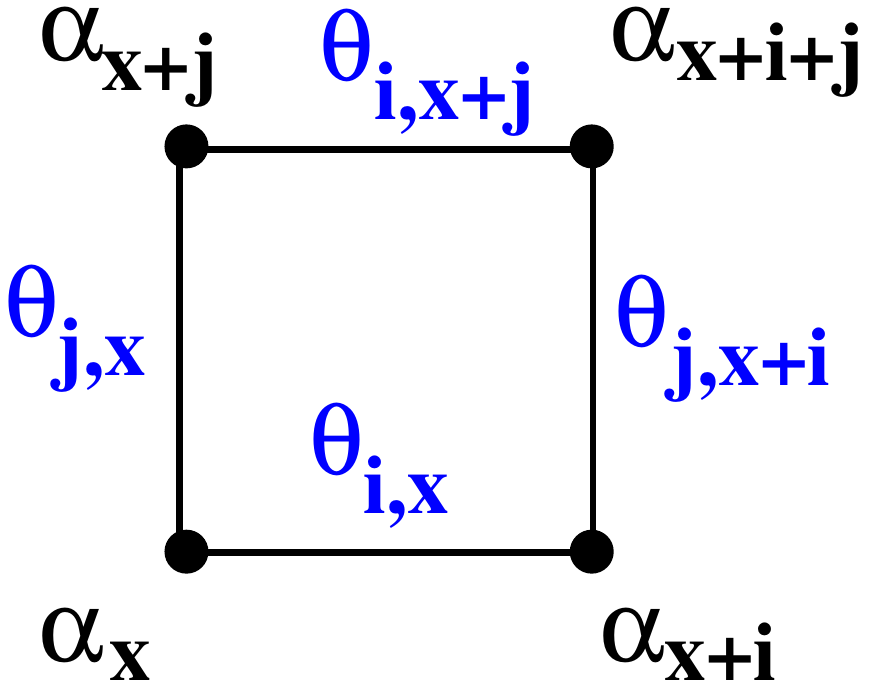}
\caption{\label{fig:Win} Variables for computing winding of a lattice plaquette. }
\end{figure}

In more detail, the winding of a plaquette is computed from  
(see Fig.~\ref{fig:Win})
\ben 
 Y_{i,\bx} = [\theta_{i,\bx} + (\al_\bx - \al_{\bx+\bi}) ]_\pi - \theta_{i,\bx}, 
\een
where 
\( \al_\bx  = [\arg \phi_\bx]_\pi \), and \(-\pi < [ \ldots ]_\pi \le \pi \).
The winding around plaquette $ij$ is then
\ben  
W_{ij,\bx} = (Y_{i,\bx} + Y_{j,\bx+i} - Y_{i,\bx+j} - Y_{j,\bx} )/2\pi.
\label{e:WinDef}
\een

 As winding has a sign and is associated with a plaquette, it can be thought of as a flux, although it is not the same as the magnetic flux.
Indeed, the magnetic flux is smoothly distributed over the width of the string, while the winding is a singular object, occupying only one lattice plaquette in a cross-section through the string. 
The centre of the strings can be defined to be the set of plaquettes with \( |W_{ij,\bx}| \ne 0 \).
Flux conservation ensures that every lattice cell has zero net winding flux, and so, plaquettes with winding can be unambiguously ordered into a closed loop. 
However, since we are on a lattice, this does not produce a smooth curve.

The positions of the strings, defined by the centres of the lattice cells whose plaquettes have winding, 
are then smoothed by averaging over nearest neighbours. 
The number of neighbours in the averaging will be denoted by the parameter $A_x$: 
this is the number of neighbours on each side that have been included into the averaging sum. 
 Thus the averaging is performed over $2A_x+1$ positions. 

Once  this procedure is performed, at every time step in the simulation,  a collection of points denoting the (averaged approximate) centres of the strings is obtained. In order to estimate the velocity of the string, one needs to compare the positions at different times, to calculate the distance that the points in the strings have traveled, and the velocity is just the distance traveled divided by the time it took them to travel that distance.

The distance a point $\bX_2$ at a time $t_2$ has travelled since time $t_1$ is determined from the nearest neighbour to $\bX_2$ in the set of points at time $t_1$. The time interval $t_2 - t_1$ is also a free parameter; we denote it by the integer $A_t = (t_2 - t_1) /\De t$, where $\De t$ is the time step in the simulation. 

Both free parameters ($A_x$ and $A_t$) have to be chosen using some physical considerations. We should aim to choose them large enough so that the metropolis effect is sufficiently reduced, but not too large so as to erase possible structure in the core of the string. 
 This is resolved by approximately $n_r =  \Swid/(a\De x)$ lattice points 
with $a$ the scale factor (remember that the physical width of the string is controlled by  $\Swid = (e\phi_0)^{-1}$, Eq.\ \ref{e:SwidDef}). We average over one string width in space,   so $A_x$ is taken to be the nearest integer to $n_r$.  

For the $s=0$ simulations this is easy to calculate, since the string width is constant in comoving coordinates during the simulation, and given that 
 $\Swid = 1$ and $\De x = 0.5$, we take 
$A_x=2$. For $A_t$ we choose the time it takes to cross the width of a string $A_t=2 A_x {\Delta x}/{\Delta t}$, where 
$\Delta t$ is the time step. In our simulations ${\Delta x}/{\Delta t}=5$, and thus, for $s=0$, we take $A_t=20$.

The parameters are more complicated in the $s=1$ simulations, because the string width shrinks as the simulation evolves. Bearing in mind how the width depends on the scale factor, and how the scale factor evolves in radiation and matter eras, one can define the ranges  for the different times in the simulation, which have been compiled in Table~\ref{ax}.

\begin{table}
\begin{tabular}{|c|c|c|c|}
 \hline
 \multicolumn{2}{|c|}{time} &$A_x$ & $A_t$ \\\hline
  Radiation & Matter &    &   \\ \hline
- & $< 663$ & 6 & 60   \\\hline
$< 488$ & [633,733]  & 5 & 50 \\ \hline 
 [488,628] & [733,831] & 4 & 40 \\ \hline
 [628,879] & [831,983] & 3 & 30\\ \hline
$>879$ &$>993$ & 2 & 20\\ \hline
\end{tabular}
 \caption{\label{ax}  Values of the parameters that determine how  velocities from the string positions  have been estimated.
}
\end{table}

\section{Validation with string standing wave}

In this section we test the length and velocity estimators on field configurations which are designed to behave like Nambu-Goto strings: small-amplitude standing waves on a NO vortex stretched around the simulation volume.
We choose the string to lie in the $(x,z)$ plane, on the curve  given by 
\ben
X = a_x  \Lbox \cos( k_z Z), \quad Y = \Lbox/2,
\label{e:StrCur}
\een
where $k_z = 2\pi/\Lbox$. 

The string is prepared by generating a series of contiguous cells $\{{\cal C}_\bx\}$ containing the curve  (\ref{e:StrCur}). The plaquettes $\{ {\cal P} \}$ separating the cells are given a flux $2\pi$ by setting their links to $\pi/2$. Other links are set to zero, and the scalar field $\phi$ is set to $\phi_0$ everywhere. The field is then cooled by gradient flow for 150, 600, or 2400 iterations for lattice spacing $\LatSpa = 0.5$, $0.25$, or $0.125$ respectively.
 
The cooling reduces the amplitude slightly, to a value $a_x^\text{eff}$ which can be estimated from the period of the oscillation. 
We choose $a_x = 0.2$ as initial conditions for the field theory simulations, which results in a field configuration shown in Fig.~\ref{f:WavIni}, along with the positions of the plaquettes $\{ {\cal P} \}$ selected by the initialisation process. The effective amplitude is $a_x^\text{eff}=0.17$ in all three cases.
Note that the relaxation also generates a Dirac string at the original position of the string (see Fig.~\ref{f:DirStr}).

\begin{figure}
   \centering
   \includegraphics[width=0.4\textwidth]{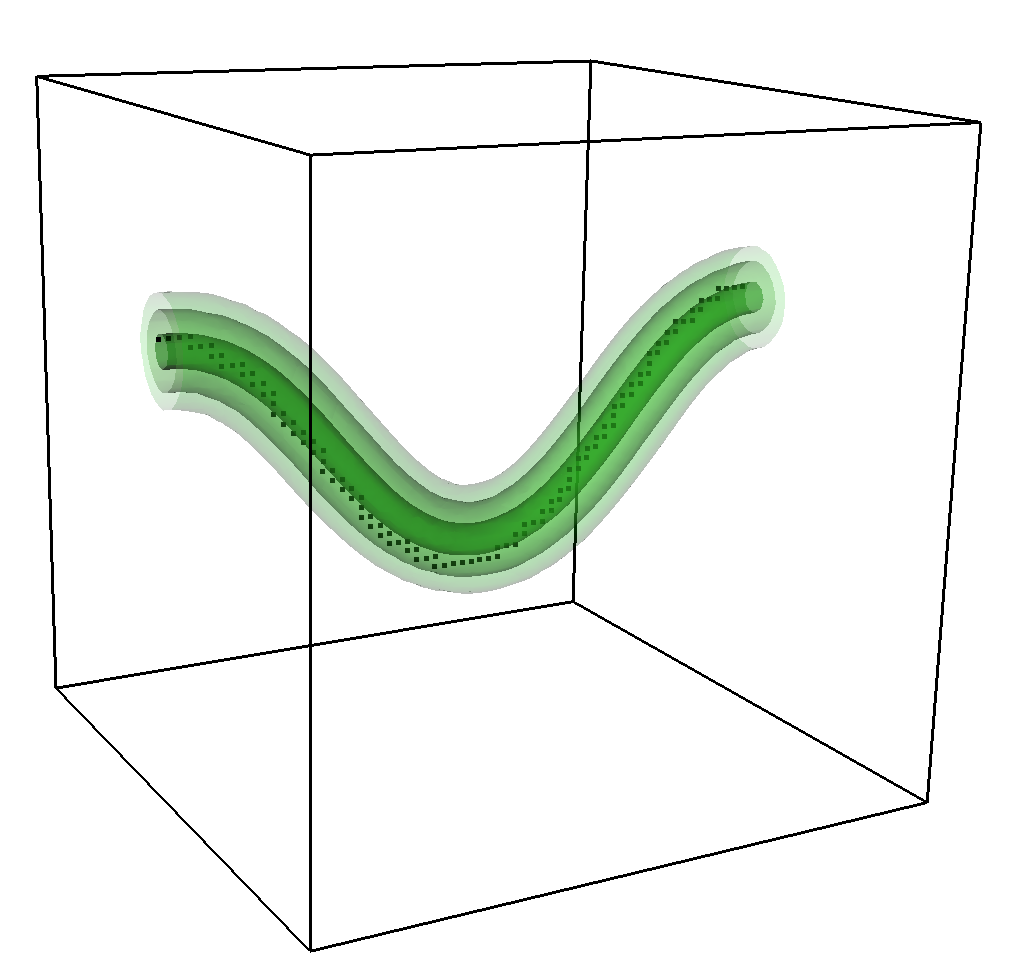} 
   \caption{Energy isosurfaces of a standing wave configuration with initial amplitude parameter $a_x = 0.2$, in a box of $N=64$ sites on a side, with lattice spacing $\LatSpa=0.5$. The dark squares show the positions of the plaquettes initialised with flux $2\pi$. The isosurfaces are at energy densities of $0.01$, $0.1$ and $1$ in units where $\phi_0=1$.}
   \label{f:WavIni}
\end{figure}

\begin{figure}
   \centering
   \includegraphics[width=0.4\textwidth]{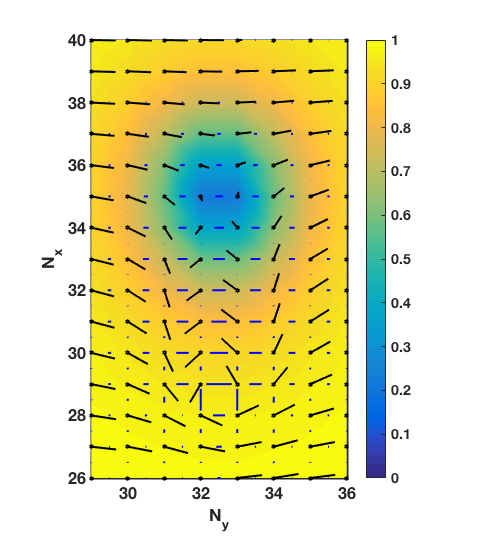} 
   \caption{Cross-section through a $64^3$ lattice containing a string standing wave, showing both the physical string and the Dirac string.
   The scalar field $\phi_\bx$ is shown as a black line at each lattice site, representing its amplitude and complex argument, 
   with blue lines joining them proportional to the value of the link variable $\theta_{i,\bx}$. 
   The modulus of the scalar field is shown as a smoothed colour field. Plaquettes with non-zero winding (see Eq.~\ref{e:WinDef}) are at $(x,y) = (34,32)$ and $(x,y) = (28,32)$. 
   The first is the physical string, the second the Dirac string.   \label{f:DirStr}}
\end{figure}

\subsection{Field estimators for string length}

\begin{figure}[h]
\centering
\includegraphics[width=.4\textwidth]{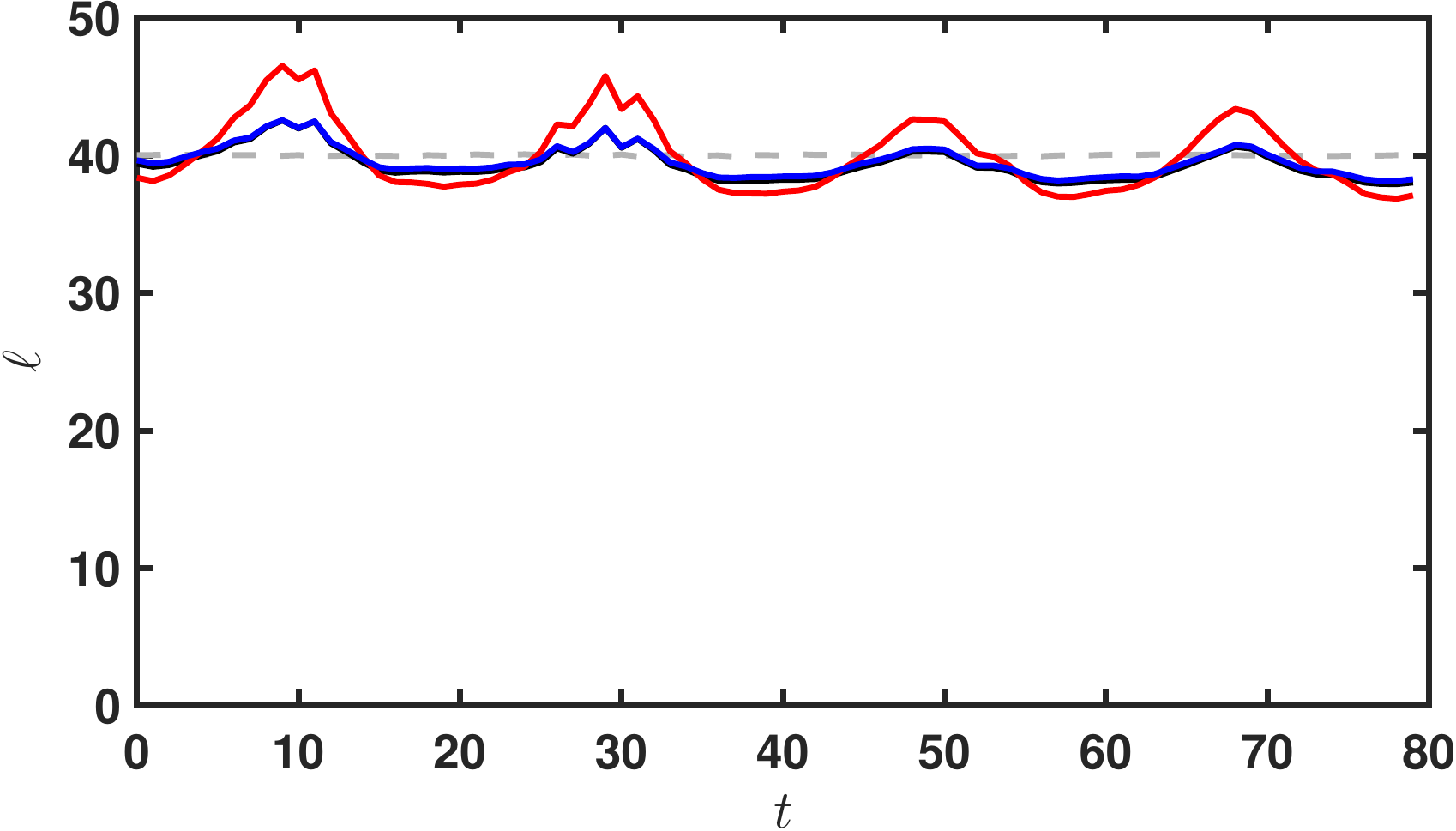}
\includegraphics[width=.4\textwidth]{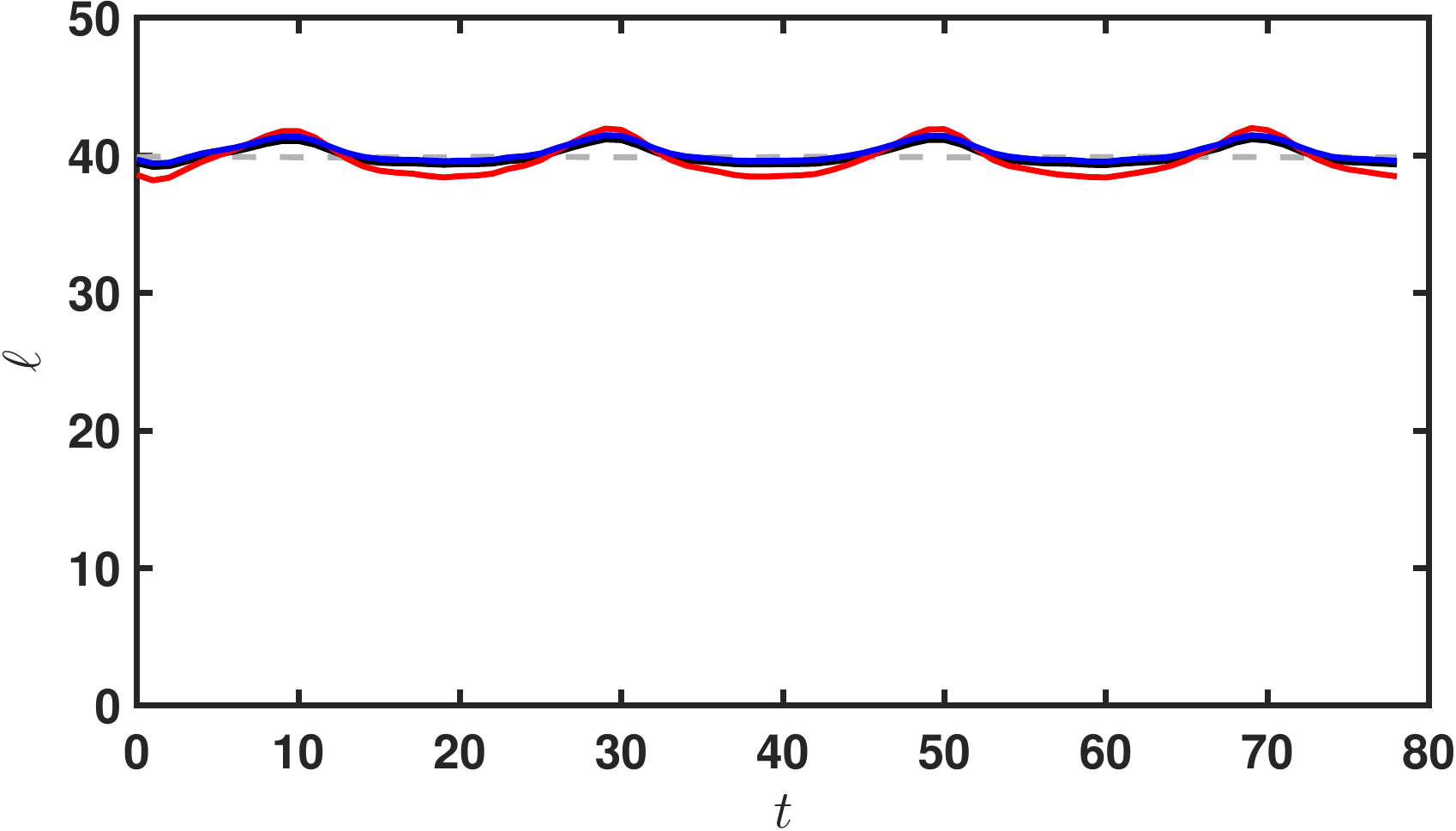}
\includegraphics[width=.4\textwidth]{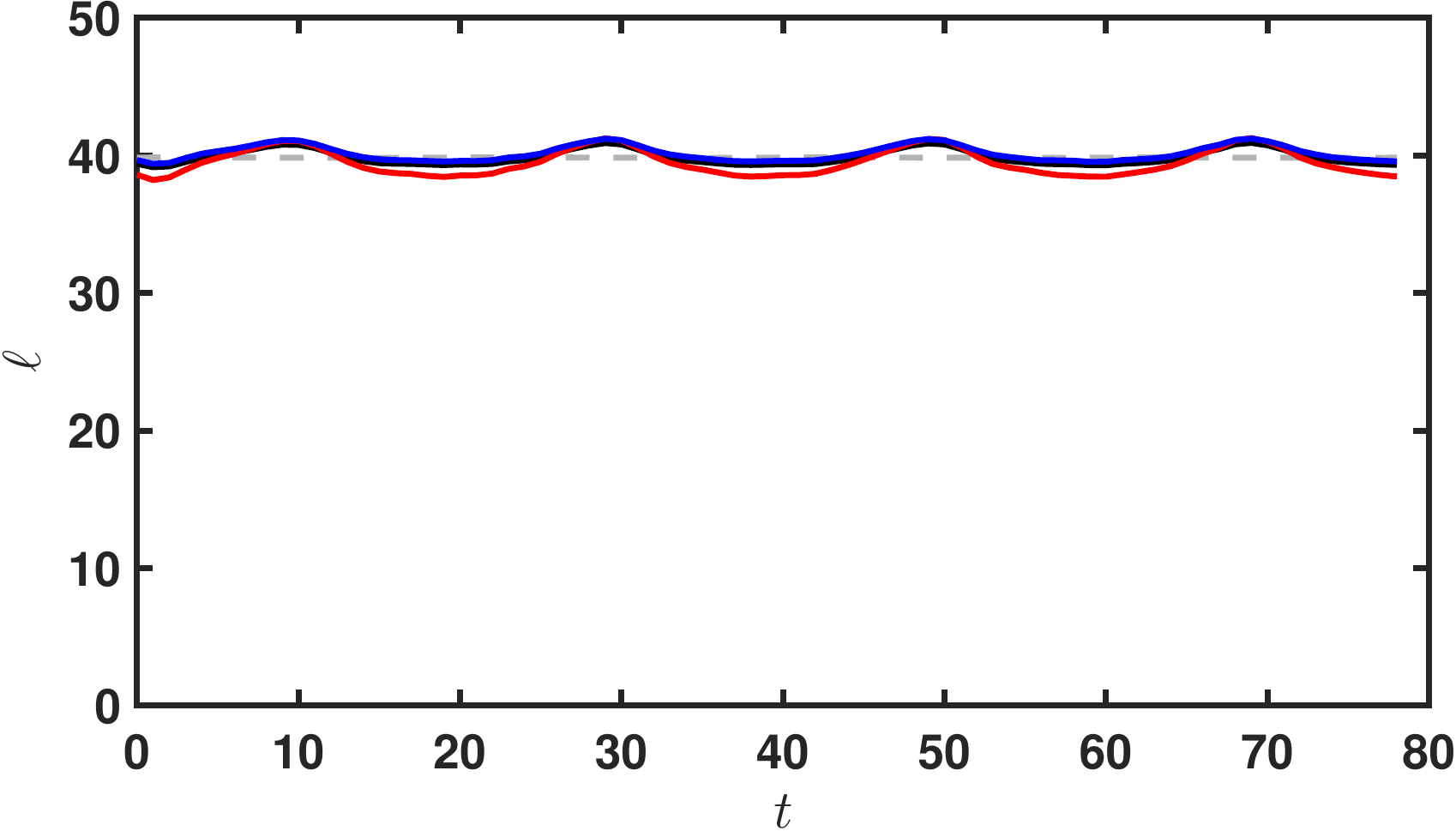}
\caption{Plots of string length estimators  
against time for a standing wave with initial amplitude parameter 
$a_x = 0.2$, in a box with $\Lbox=32$, with lattice spacing (from top to bottom) 
$\Delta x = 0.5$, $\Delta x = 0.25$  and $\Delta x = 0.125$.
The length estimators are 
$\ell_{\mcL}$ (black, Eq.~\ref{e:ell_lag}),
$\ell_{B,\mcL}$ (red, Eq.~\ref{e:ell_B_lag}), and
$\ell_{s,\mcL}$ (blue, Eq.~\ref{e:ell_s_lag}), The length estimators and simulation times are given in units where $\phi_0=1$.
The total energy is shown in dashed grey.
}
\label{fig:ell_wave}
\end{figure}

In Fig.~\ref{fig:ell_wave} we show graphs of the string length estimators (\ref{e:ell_lag}), (\ref{e:ell_B_lag}) and (\ref{e:ell_s_lag}) for the standing wave configuration described above, in a box of physical size $\Lbox=32$, with lattice spacings $\Delta x = 0.5, 0.25$, and $0.125$ (top to bottom).
The length estimators are close to constant, and so are good estimates of the invariant string length, although the coarsest lattice spacing the magnetic field estimator  (red, Eq.~\ref{e:ell_B_lag}) is more than 10\% above the others. 
For the other estimators the departure from a constant, about 5\% when the string speed is highest, is very similar for  $\Delta x = 0.25$ and $0.125$, demonstrating that it is not a lattice effect.  It is about the same order of magnitude as the ratio thickness of the string to the local curvature radius.

\subsection{Estimators for string velocity}
\label{ss:V2_wave}

In this section we show the results for the mean square velocities from the standing wave described above.
In Fig.~\ref{fig:V2_wave} we show graphs of the string velocity estimators (\ref{e:v2_lag}), (\ref{e:v2_B_lag}), (\ref{e:v2_s_lag}) and (\ref{e:v2_w_lag}) for a standing wave configuration on lattices with different spatial resolutions.
The length estimators are close to the value predicted for a Nambu-Goto string of the same length. As with the length estimators, the largest departures for the magnetic field estimator at the coarsest lattice spacing.
The best appears to be the equation of state estimator $\vAv^2_{w,\mcL}$ (green, Eq.~\ref{e:v2_w_lag}).
The fact that this estimator is slightly lower on the second cycle is a sign that a small amount of energy is being lost to lattice radiation.

In order to estimate the velocities from the positions of the plaquettes with winding, the Dirac string needs to be removed.
To do this
we compare the simulations at three different times: when the oscillating string is straight, when it is at its maximum excursion in phase with the Dirac string, and when it is at its maximum excursion out phase with the Dirac string. The points that are common to those three configurations are removed.

Note that, inevitably, there are points shared by the Dirac string and the oscillating string which are also removed. Thus, we do not recover a connected string, but only segments of it. 
The missing segments are generally short and are filled in by linear interpolation if the gaps are too large to be covered by the smoothing window.

The resulting mean squared velocity estimates are shown in Fig.~\ref{fig:V2_wave} as black circles. 
We see that the velocity estimates from the positions of plaquettes with winding fluctuate around the peaks, 
and are systematically lower than the true value by about 15\%.  This observation is consistent with the findings of Ref.~\cite{Moore:2001px}. 

We have experimented with applying the algorithm directly to the Nambu-Goto solution evaluated at the same times as the snapshots from the numerical simulations. We find that the $\bar{v}^2$ is accurately reproduced when the floating point values are used, but similar fluctuations and underestimates are found when the string positions are rounded to a lattice. We conclude that both effects are a result of the lattice discretisation.

\begin{figure}[h]
\centering
\includegraphics[width=.4\textwidth]{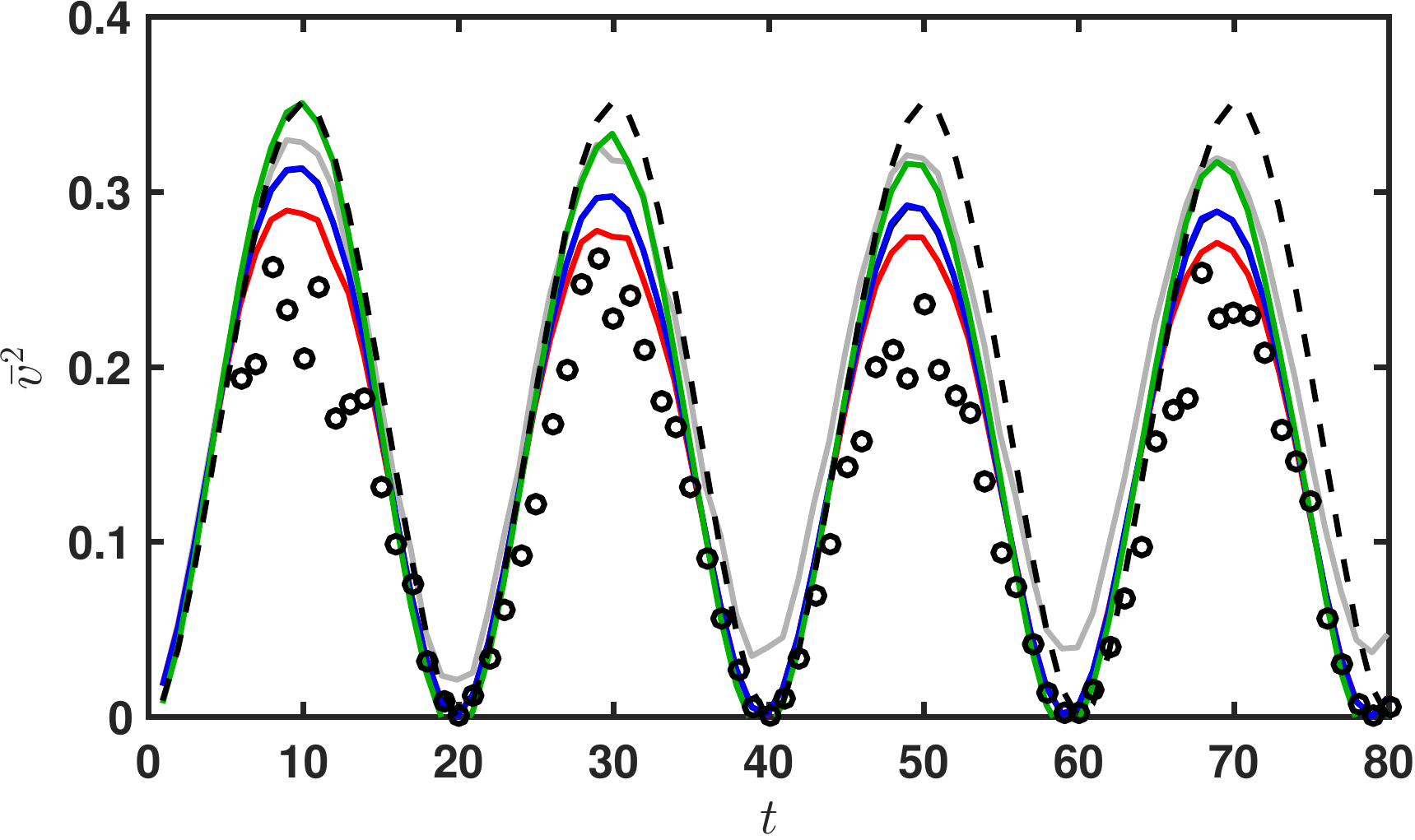}
\includegraphics[width=.4\textwidth]{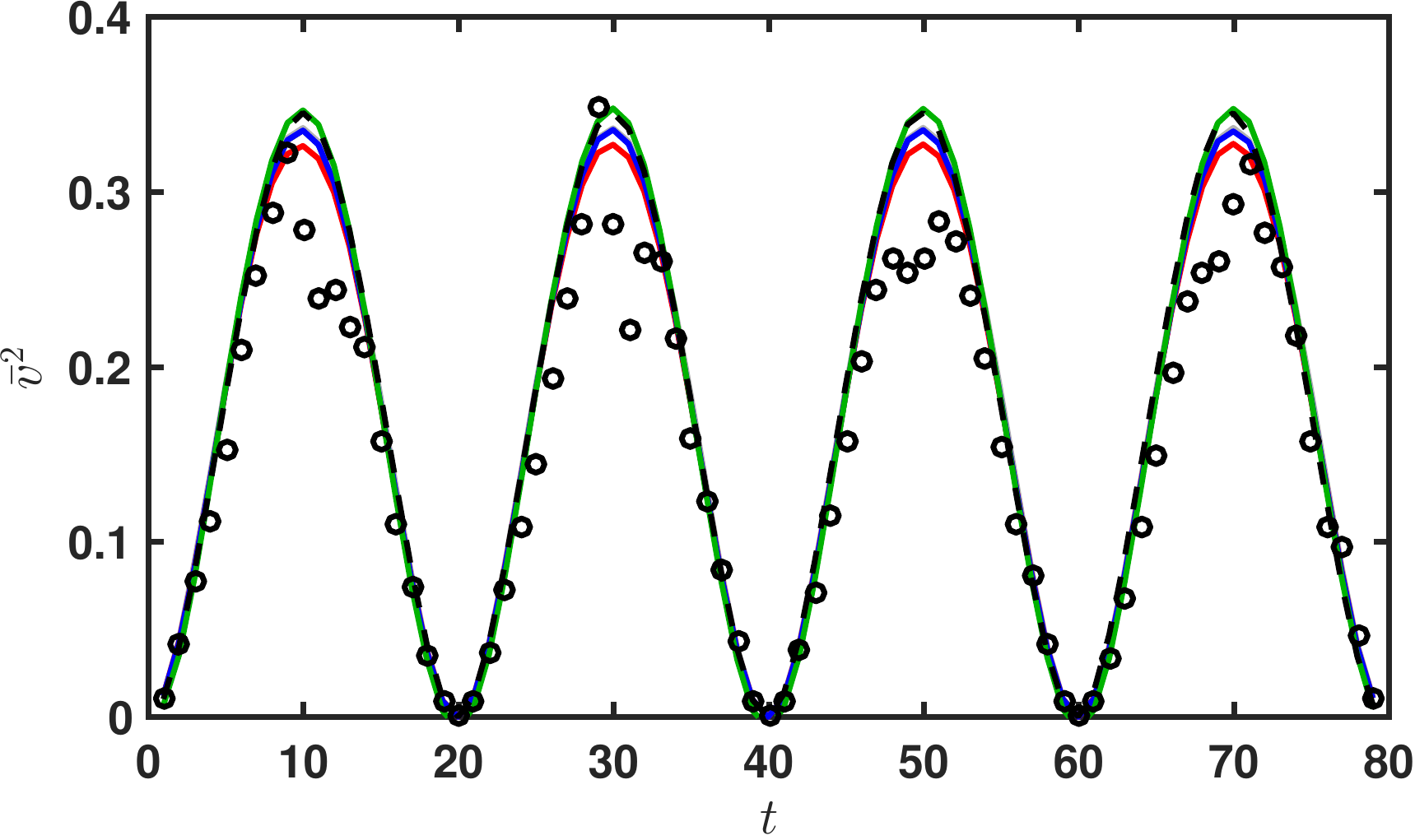}
\includegraphics[width=.4\textwidth]{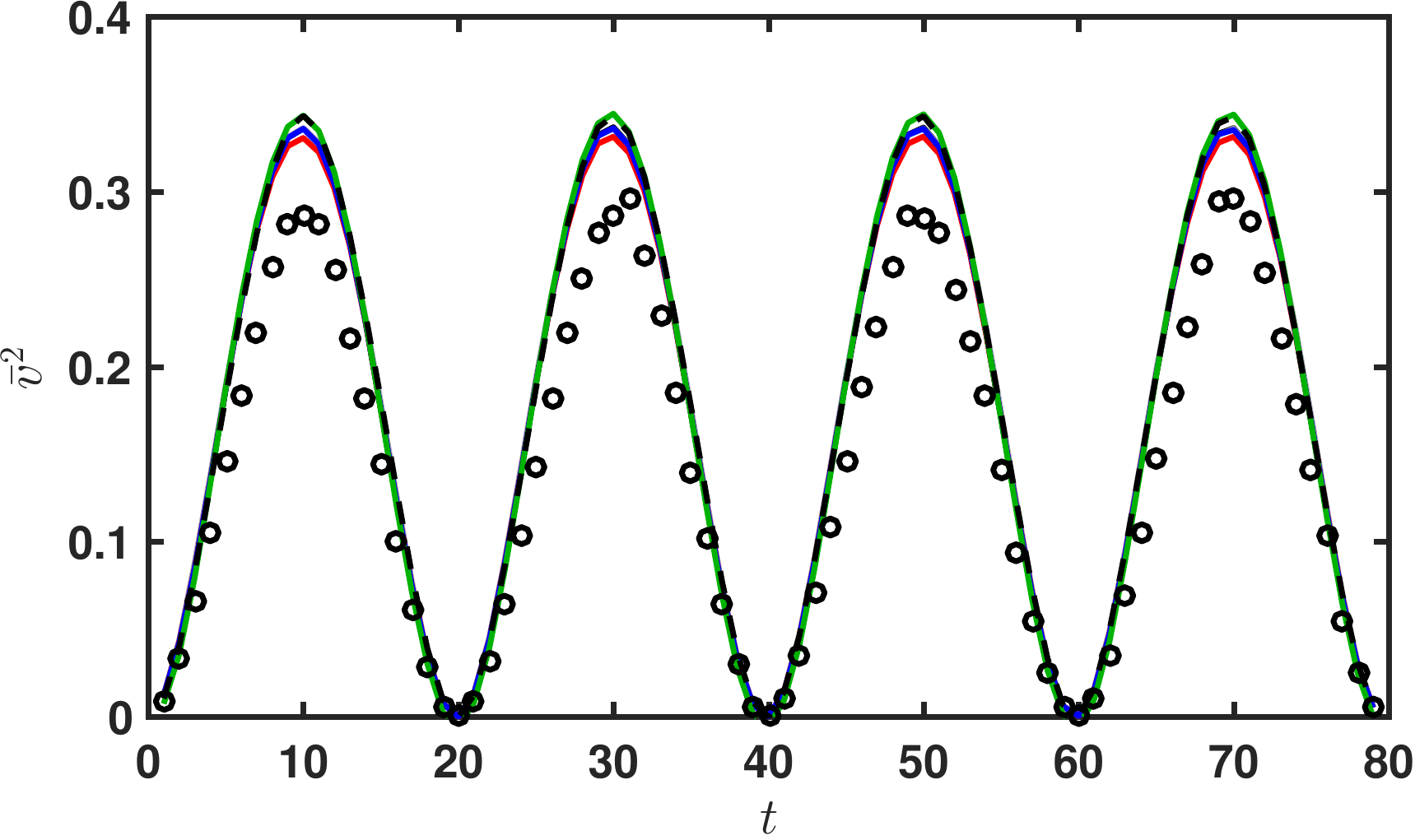}
\caption{Plots of mean square velocity estimators against time for a standing wave in a box of size $\Lbox=32$, with initial amplitude parameter 
$a_x = 0.2$, and lattice spacing 
$\Delta x = 0.5$ (top), $\Delta x = 0.25$ (centre) and $\Delta x = 0.125$ (bottom).
The velocity estimators are 
$\vAv^2_{\mcL}$ (black, Eq.~\ref{e:v2_lag}),
$\vAv^2_{B,\mcL}$ (red, Eq.~\ref{e:v2_B_lag}),
$\vAv^2_{s,\mcL}$ (blue, Eq.~\ref{e:v2_s_lag}), and 
$\vAv^2_{w,\mcL}$ (green, Eq.~\ref{e:v2_w_lag}).
The velocities estimated from the string position as given from the winding (see Sec.~\ref{velfrompos}) is shown with	 open circles. 
The Nambu-Goto prediction is shown in dashed black.
Note that $\vAv^2_{\mcL}$ and $\vAv^2_{s,\mcL}$ are almost identical. The lengths and  times are given in units where $\phi_0=1$.
}
\label{fig:V2_wave}
\end{figure}

\section{String networks}

The previous section tested the accuracy and reliability of the estimators for the string length 
and the mean square velocity from local quantities built from the fields of the Abelian Higgs model.  
We concluded that the most reliable were the length estimate 
$\Slen_\mcL$ (\ref{e:ell_lag}) and the mean squared velocity estimator $\vAv^2_{w,\mcL}$ (\ref{e:v2_w_lag}). 
In this section we apply them to string network simulations in an expanding space-time background.

We performed simulations on comoving cubic lattices of side $N=512, 1024, 2048$ and $4096$ with lattice spacing $\LatSpa = 0.5$, in matter and radiation eras, and with core growth parameter $s=1$ and $s=0$. As mentioned before, all simulations were performed at critical coupling $\lambda = 2e^2$, or in other words, $\beta=1$.
The $N=4096$ (``4k'') simulations have been described in detail elsewhere \cite{Daverio:2015nva}. 
We performed new simulations on smaller lattices in order to test the dependence of the results on the size of the lattice.

At $s=1$ the string width (\ref{e:SwidDef})  
shrinks from a maximum at $\tCG$, to 1 at the end of the simulation, $\tEnd$.  
It is important to realise that this means that $s=1$ simulations sample the core of the string with a greater number of points than $s=0$ simulations for most of the period over which data is taken, and thus are also tests of resolution effects. 

Table \ref{t:simpars} gives the important simulation parameters, including the time ranges over which the principal data are extracted, 
divided into ``early'' and ``late''. The early time range has had less time to evolve towards scaling, but resolves the string better in the $s=1$ 
simulations.

We will sometimes fit separately on the $s=0$ and $s=1$ datasets,  
in which case we will refer to them as s0, s1. 
Sometimes we will distinguish between the data in the early and late time ranges, in which case we use the letter code e and l.  
So, for example, s1l denotes the data from the 22 simulations with $s=1$ in the larger of the two time ranges.

We also performed a set of simulations with $N=4096, 2048$ and $1024$ at resolutions $\LatSpa=0.125, 0.25$ and $0.5$, respectively, in order to test the resolution.  The smaller lattice initial conditions were decimations of the $N=4096$ simulation after the initial conditions are prepared. 
Parameters are given in Table \ref{t:simpars_dx}, and the dataset letter code is R.

From the simulations we extract the total length of string $\ell$ using the Lagrangian-weighted energy,
and the mean square velocity $\vAv^2$ using the Lagrangian-weighted equation of state.
Rather than using the total length of string $\ell$, we define the mean comoving string separation 
\ben
\xi = \sqrt{\frac{\vol}{\ell}},
\een
which should increase linearly with conformal time $\ta$ in a scaling network, 
and so $\dot\xi $ should tend to a constant.

The important quantities $\dot\xi$ and $\vrms^2$ are estimated in two different time ranges over the simulation, 
the first with a linear fit to $\xi$, and the second from an average over the $\vAv^2_{w,\mcL}$ 
estimator recorded in the range. 
The uncertainty in $\dot\xi$ is dominated by the standard deviation in slopes between simulations.  
The uncertainties in $\vAv^2$ are calculated from combining in quadrature the variance of $\vAv^2$ over the time range and 
the variance in the mean between simulations.

\begin{table}
\renewcommand{\arraystretch}{1.2}
\begin{tabular}{|c||c|c||c|c|}
 \hline
Lattice size & \multicolumn{4}{c|}{$N=512$} \\\hline
 Core growth & \multicolumn{2}{c||}{$s=1$} & \multicolumn{2}{c|}{$s=0$}  \\\hline
 Cosmology & Radiation & Matter & Radiation & Matter \\ \hline
 $\tau_\text{cg}$ & 60 & 75  & -- & -- \\
 $\Swid^\text{max}$ &  2.5 & 4.0 & 1 & 1 \\
 $\tau_{\mathrm{end}}$ & 150 & 150 & 150 & 150 \\
 $\tau \in$ (early) & [100,125]  & [100,125] & [100,125]  & [100,125] \\
 $\tau \in$ (late) &   [125,150] & [125,150] & [125,150] & [125,150]\\
 $n_\text{sim}$ & 5 & 5 & 5 & 5 \\
 \hline
 \hline
Lattice size & \multicolumn{4}{c|}{$N=1024$} \\\hline
  Core growth & \multicolumn{2}{c||}{$s=1$} & \multicolumn{2}{c|}{$s=0$}  \\\hline
 Cosmology & Radiation & Matter & Radiation & Matter \\ \hline
 $\tau_\text{cg}$ & 100 & 150  & -- & -- \\
 $\Swid^\text{max}$ & 3.0 & 4.0 & 1 & 1 \\
 $\tau_{\mathrm{end}}$ & 300 & 300 & 300 & 300 \\
 $\tau \in$ (early)  &  [150,200] & [175,225] &  [150,200] & [150,200] \\
 $\tau \in$ (late) &  [250,300] &  [250,300] &  [250,300] & [250,300] \\
 $n_\text{sim}$ & 5 & 5 & 5 & 5 \\
 \hline
 \hline
Lattice size & \multicolumn{4}{c|}{$N=2048$} \\\hline
  Core growth & \multicolumn{2}{c||}{$s=1$} & \multicolumn{2}{c|}{$s=0$}  \\\hline
 Cosmology & Radiation & Matter & Radiation & Matter \\ \hline
 $\tau_\text{cg}$ & 150 & 245 & -- & -- \\
 $\Swid^\text{max}$ & 4.0 & 6.0 & 1 & 1 \\
 $\tau_{\mathrm{end}}$ & 600 & 600 & 600 & 600 \\
 $\tau \in$ (early)  & [300,350] & [400,450] & [300,350] & [300,350] \\
 $\tau \in$ (late) & [550,600] & [550,600] & [550,600] & [550,600] \\
 $n_\text{sim}$ &6 &6 & 6 &  6\\
 \hline
\hline
Lattice size & \multicolumn{4}{c|}{$N=4096$} \\\hline
  Core growth & \multicolumn{2}{c||}{$s=1$} & \multicolumn{2}{c|}{$s=0$}  \\\hline
 Cosmology & Radiation & Matter & Radiation & Matter \\ \hline
 $\tau_\text{cg}$ & 204 & 366 & -- & -- \\
 $\Swid^\text{max}$ & 5.4 & 9.0 & 1 & 1 \\
 $\tau_{\mathrm{end}}$ & 1100 & 1100 & 1100 & 1100 \\
 $\tau \in$ (early)  & [600,650] & [600,650] & [600,650] & [600,650] \\
 $\tau \in$ (late) & [1050,1100] & [1050,1100]  & [1050,1100] & [1050,1100] \\
 $n_\text{sim}$ & 6 & 6 & 7 & $7^*$ \\
  \hline
\end{tabular}
 \caption{\label{t:simpars} 
Simulation parameters for the data shown in Figs.\  \ref{fig:xidot_lag}, \ref{fig:v2_dx} and \ref{fig:RadEffParDx},
from which the s0 and s1 datasets are derived.
Given are:
conformal time at end of core growth $\tau_\text{cg}$,
the maximum comoving width of the string $\Swid^\text{max}$  (i.e. the string width at  $\tau_\text{cg}$),
conformal time at end of simulation $\tau_{\mathrm{end}}$,
and the two conformal time ranges over which the rate of change of the string separation parameter $\dot\xi$ and the mean square velocity $\bar{v}^2$ are obtained. The parameter $n_\text{sim}$ denotes how many simulations of each type were performed. 
Note that mean square velocity data was obtained for only one of the seven $s=0$ matter era runs.
}
\end{table}

\begin{table}
\renewcommand{\arraystretch}{1.2}
\begin{tabular}{|c||c|c|}
 \hline
Lattice size & \multicolumn{2}{c|}{$N=1024$} \\\hline
 Cosmology & Radiation & Matter \\ \hline
 $\tau_{\mathrm{end}}$ & 300 & 300 \\
 $\tau \in$ (early) &  [150,200] & [150,200] \\
 $\tau \in$ (late) &  [250,300] & [250,300] \\
 \hline
 \hline
Lattice size & \multicolumn{2}{c|}{$N=2048$} \\\hline
 Cosmology & Radiation & Matter \\ \hline
$\tau_{\mathrm{end}}$ & 600 & 600 \\
 $\tau \in$ (early) & [300,350] & [300,350] \\
 $\tau \in$ (late) & [550,600] & [550,600] \\
 \hline
\hline
Lattice size & \multicolumn{2}{c|}{$N=4096$} \\\hline
 Cosmology & Radiation & Matter \\ \hline
 $\tau_{\mathrm{end}}$ & 1100 & 1100 \\
 $\tau \in$ (early) & [600,650] & [600,650] \\
 $\tau \in$ (late) & [1050,1100] & [1050,1100] \\
  \hline
\end{tabular}
 \caption{\label{t:simpars_dx} 
Simulation parameters for the resolution test run from which the R dataset is derived.
Given are:
the comoving width of the string $\Swid^\text{max}$,
conformal time at end of simulation $\tau_{\mathrm{end}}$,
and the two conformal time ranges over which the rate of chance of the string 
separation parameter $\dot\xi$ and the mean square velocity $\bar{v}^2$ are obtained. 
All simulations use $s=0$ core growth, with constant comoving string width $\Swid = 1/\phi_0$.
Only one simulation was performed at each resolution.
}
\end{table}

\subsection{String length estimators}

In Fig.~\ref{fig:xi_lag} we show the Lagrangian-weighted energy estimate of the string separation parameter $\xi$ as a function of conformal time $\tau$ for  the simulations listed in Table \ref{t:simpars}.

\begin{figure}[h]
\centering
\includegraphics[width=.4\textwidth]{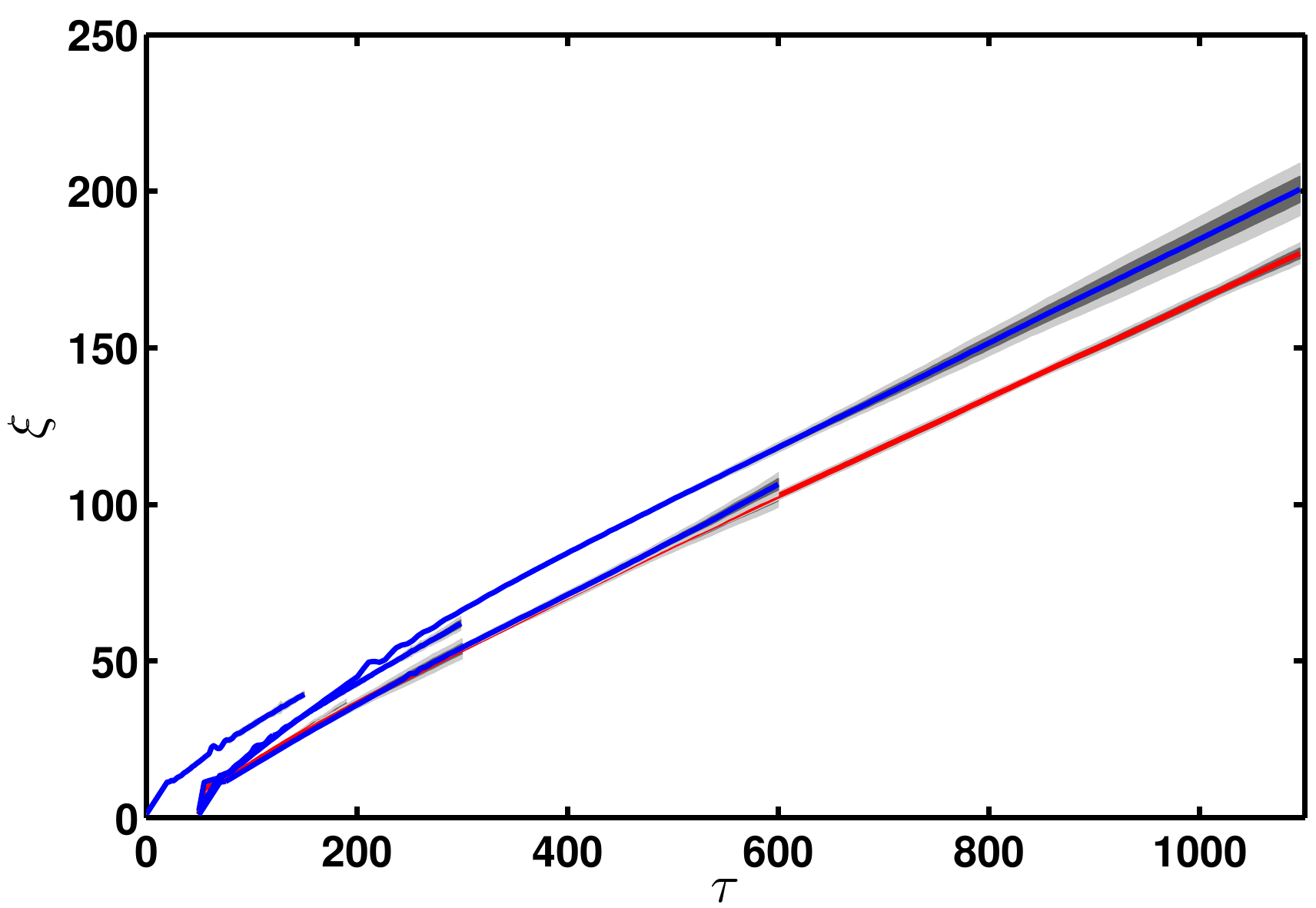}\\
\includegraphics[width=.4\textwidth]{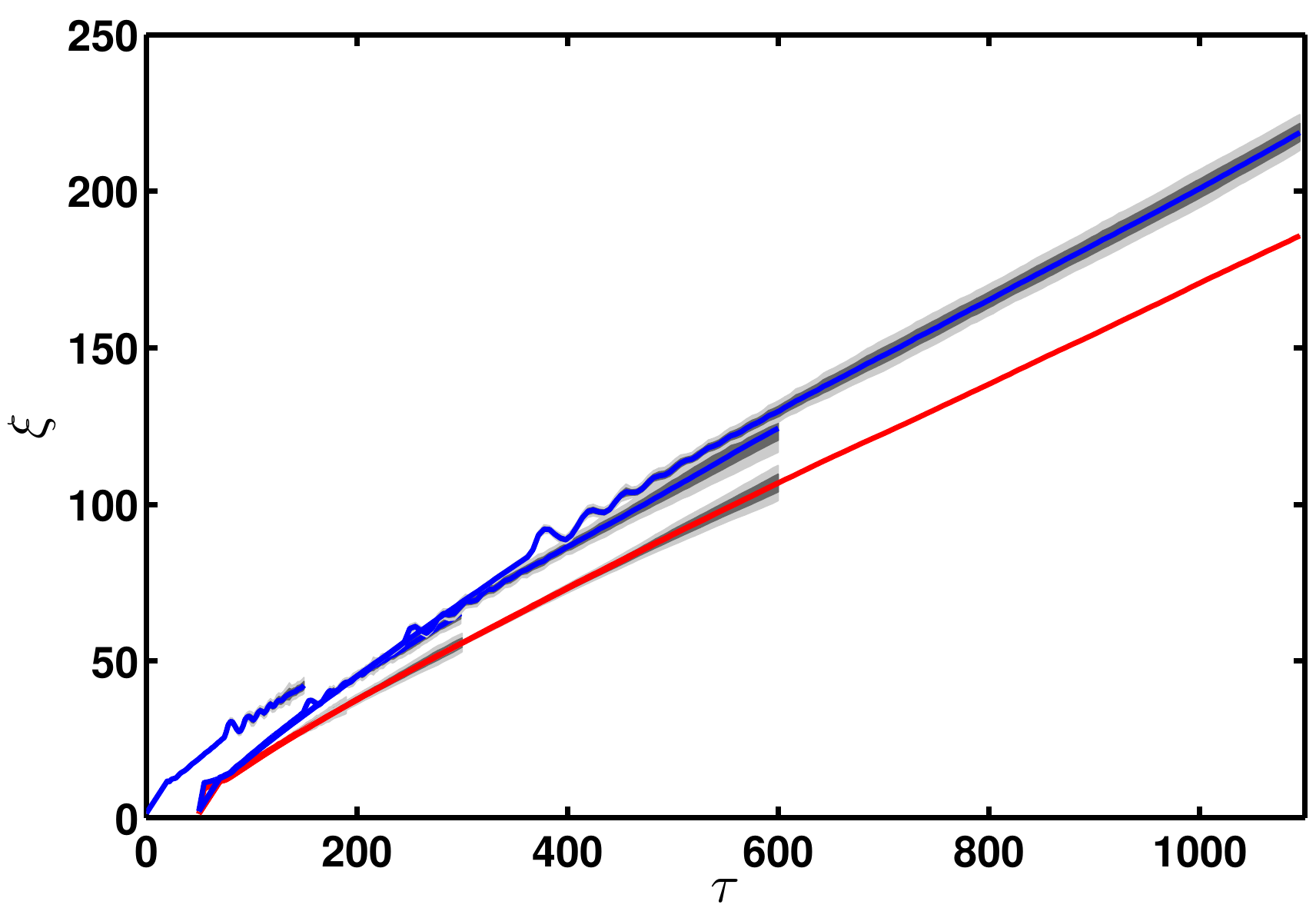}
\caption{Plots of $\xi_\mcL$ versus time for all network simulations in the radiation (top) and matter (bottom) eras. The error bands mark the 
	1-$\si$ and 2-$\si$ variations between realisations.
	Simulations with $s=0$ are marked in red, $s=1$ in blue.  The string separation parameter and simulation times are given in units where $\phi_0=1$.
	}
\label{fig:xi_lag}
\end{figure}

The general impression is that all are tending to linear growth, independent of the lattice size and lattice spacing, after a period of relaxation from the almost stationary network generated as initial conditions. This means that Abelian Higgs string networks scale, over a range of ratios of string separation to string width $\Swid/\xi$ from $0.05$ to about $0.003$.  

We first consider the effect of lattice spacing on the slope $\dot\xi$, which is directly probed by the resolution test simulations. 
The values of $\dot\xi$ are very close: they differ by less than $0.01$ between the simulations at $\LatSpa/\Swid = 0.5$, $0.25$ and $0.125$. 
A linear least squares to the combination of the resolution test run and the $s=1$ datasets (R and s, which also explore different $\LatSpa/\Swid$)
shows no significant lattice spacing dependence (see Table \ref{t:slopes_dx}).
This is good evidence that even $\LatSpa/\Swid = 0.5$ has sufficient lattice resolution to give reliable results for $\xi$. 

We need to extrapolate $\Swid/\xi$ to a very small number to reach values relevant for CMB calculations. 
In Fig.~\ref{fig:xidot_lag} we plot  $\dot\xi$
against the ratio of the string width to the string separation $\Swid/\xi$, 
with error bars showing the fluctuations between simulations.
We also show linear fits, weighted by the error bars,
using either $s=1$ data only (dashed blue line) or $s=0$ data only (dashed red line),
both taken in the late time range so that effects from the initial conditions are minimised.
The slopes of the fit lines for $\dot\xi$ are given in Table \ref{t:slopes_xi}.
There is a clear dependence of $\dot\xi$ on the mean separation. 
The insignificant lattice spacing dependence allows us  
to take the large string separation limit as the intercept with the $y$ axis; the values  
are listed in Table \ref{t:final_fits}. 

Fig.~\ref{fig:xidot_lag} shows that the values of $\dot\xi$ observed in the $s=0$ simulations are consistently lower than in the $s=1$ simulations; 
the matter era asymptote is almost 30\% lower in the matter era.
Note also that the central values in the $s=1$ 4k simulations (furthest to the left on these graphs) are within the errors of the extrapolated values. The value of $\dot\xi$ is important for the overall normalisation of the UETCs \cite{Daverio:2015nva}. The UETC normalisation used for the calculation of the CMB power spectrum in \cite{Lizarraga:2016onn} made no use of extrapolation, but we conclude that it is consistent with the linearly extrapolated value of $\dot\xi$.

\begin{figure}[h]
\centering
\includegraphics[width=.4\textwidth]{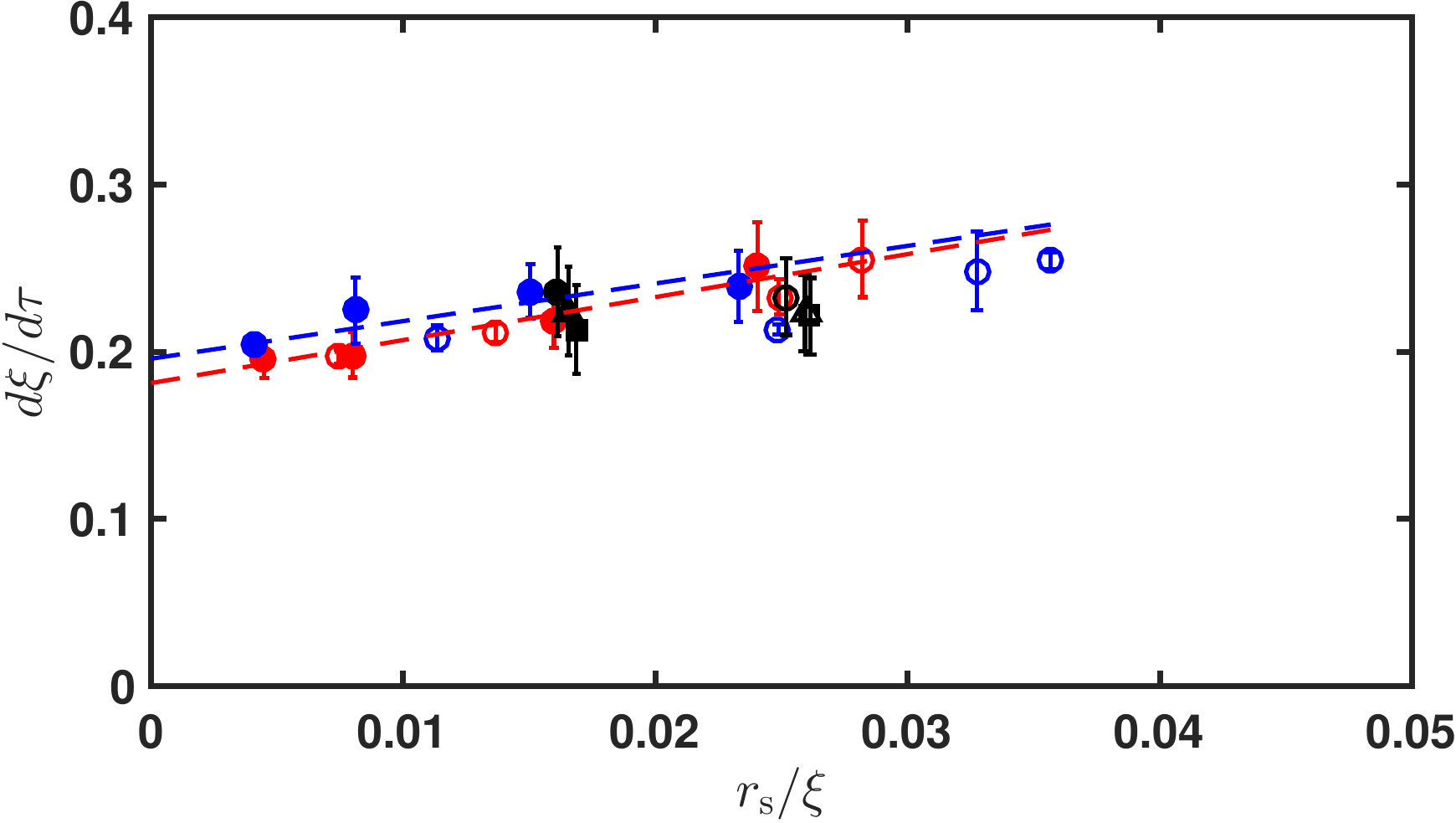}
\includegraphics[width=.4\textwidth]{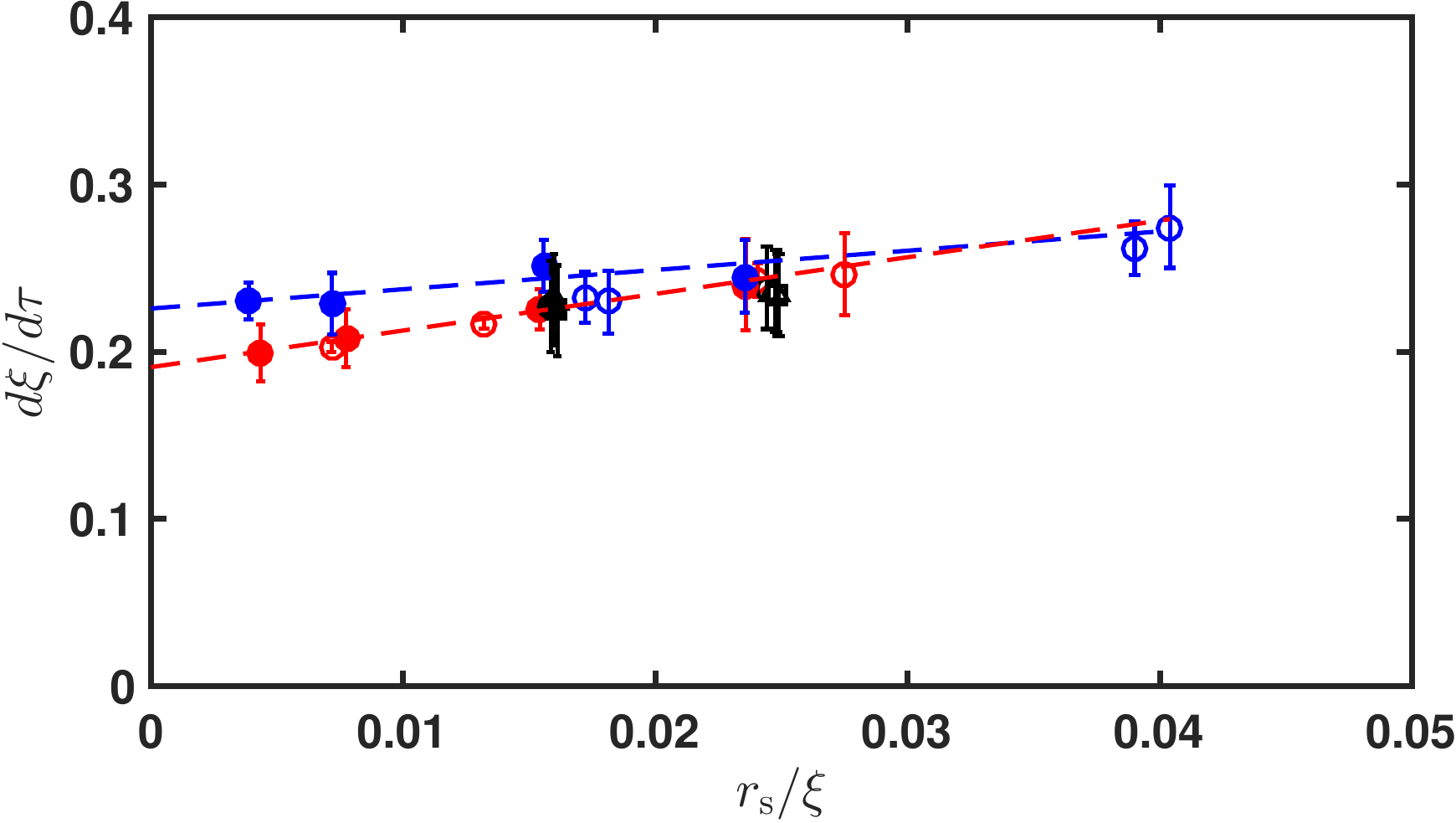}
\caption{Plots of $\dot\xi$ versus $\Swid/\xi$ for all simulations, for radiation (top) and matter era (bottom) simulations.
Two time ranges are plotted, detailed in Table \ref{t:simpars}, 
one around half way through the simulation (open symbols), and one near the end (filled symbols).
Circles in blue are $s=1$, while circles in red have $s=0$. 
Black symbols are the resolution test runs with lattice spacing $0.5$ (circle), $0.25$ (triangle) and $0.125$ (square).
Dashed lines show linear fits to the $s=1$ data and $s=0$ data separately, with the same colour code. 
}
\label{fig:xidot_lag}
\end{figure}

\subsection{String velocity estimators}

In Section \ref{ss:V2_wave} we identified the best local velocity estimator as  
$\vAv^2_{w,\mcL}$, constructed from the Lagrangian weighted equation of state (\ref{e:v2_w_lag}). 
We also found that the velocity estimates were inaccurate at lattice spacing $\LatSpa/\Swid = 0.5$. 

In Fig.~\ref{fig:V2_lag} we show the evolution of the mean square velocity for our network simulations.
Velocities calculated from the estimator  (\ref{e:v2_w_lag}) are shown as solid lines, 
while those
calculated from the string positions are shown with dashed lines. 
As with the standing wave, they are consistently lower than the field estimator by about 15\%.
The validation of the field estimator against the standing wave configuration, where the string trajectory is known, gives confidence that the field estimator is the more reliable one.

\begin{figure}[h]
\centering
\includegraphics[width=.49\textwidth]{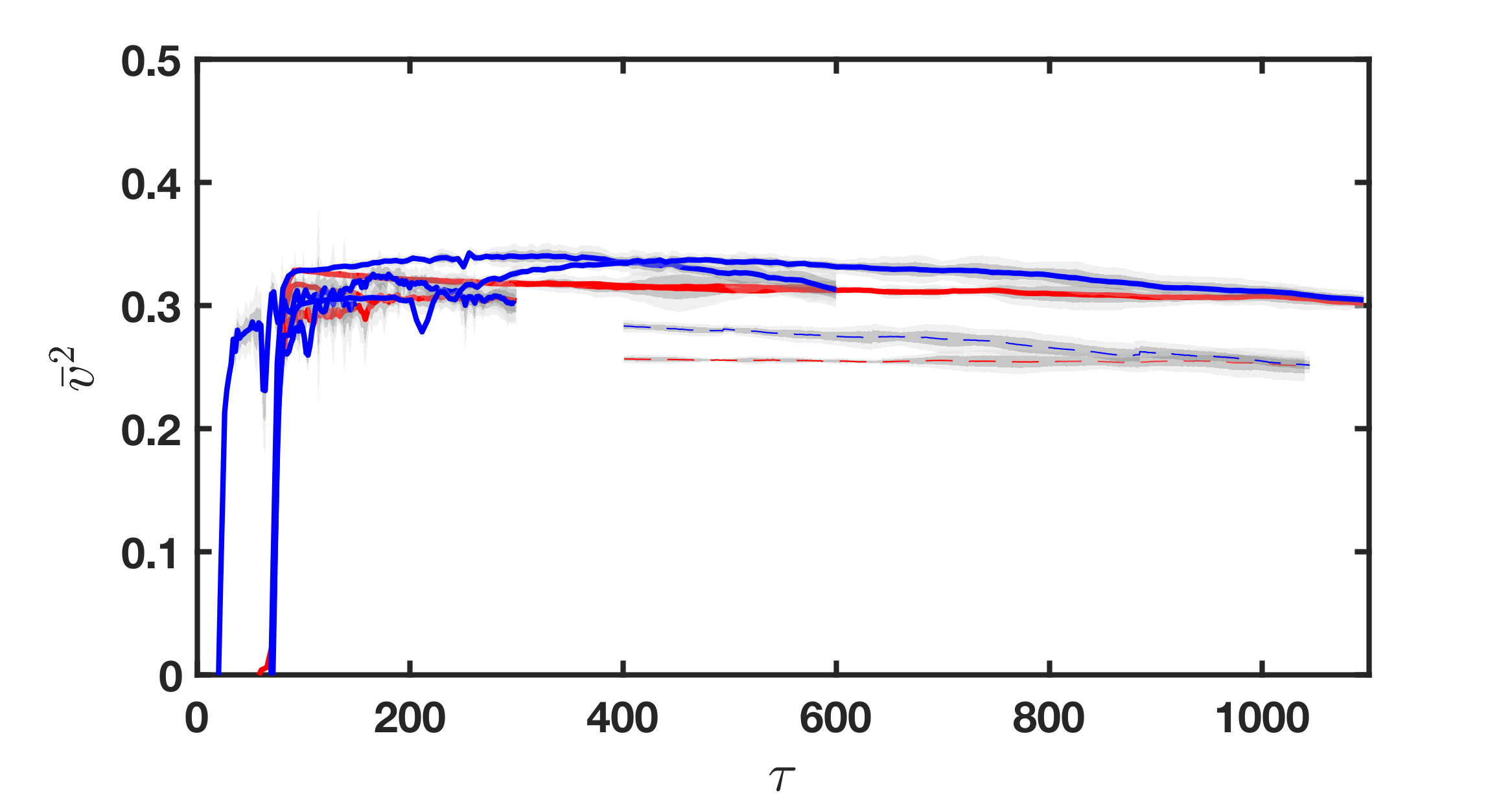}\\
\includegraphics[width=.49\textwidth]{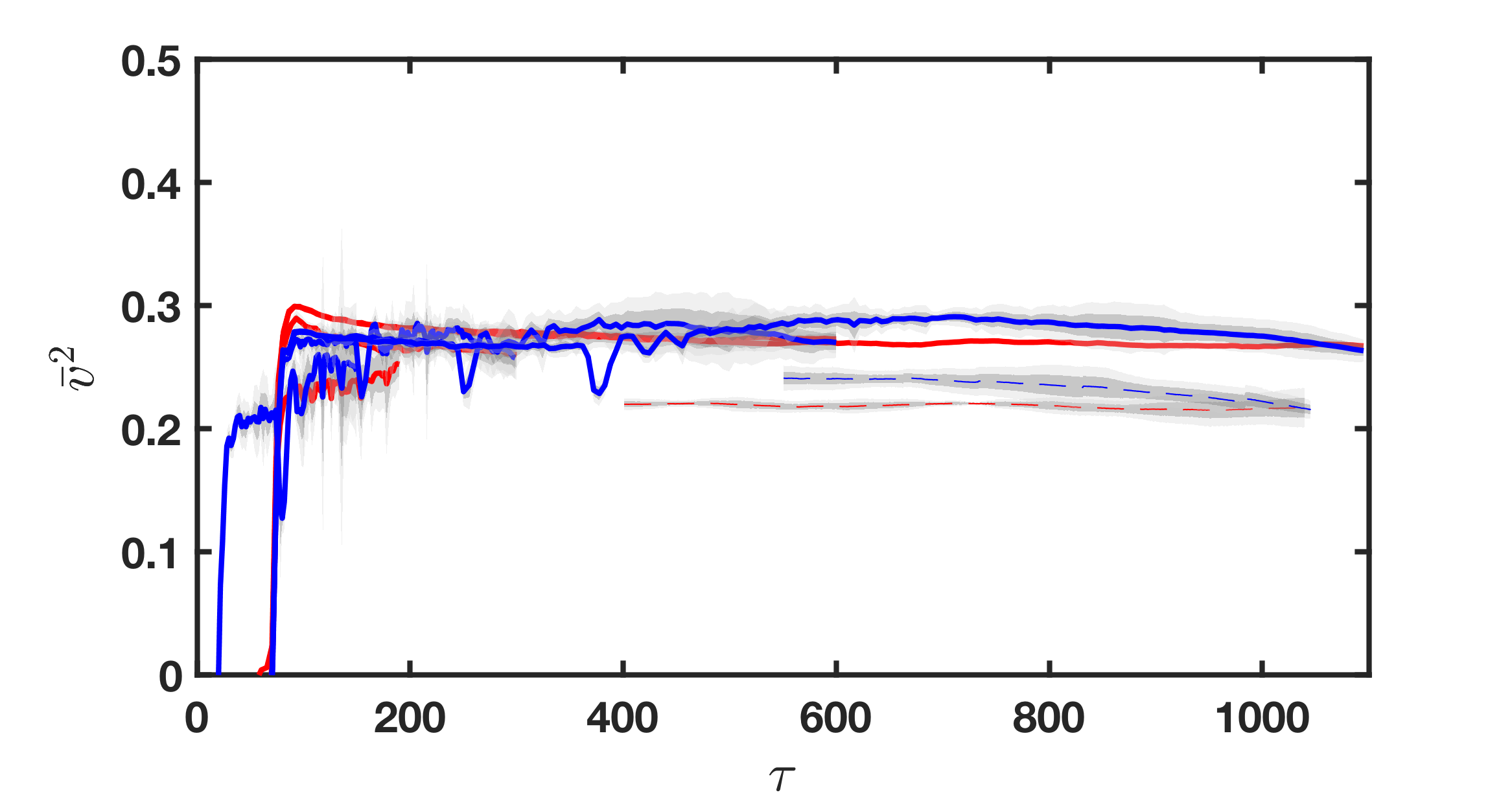}
\caption{Plots of mean square velocity $\vAv^2$ versus time for all network simulations in the radiation (top) and matter (bottom) eras.
	Simulations with $s=0$ are marked in red, $s=1$ in blue.
	Solid lines are the field estimator $\vAv^2_\mcL$, while velocities from string positions are marked with dashed lines.
	Simulations with lattice sizes $512^3$, $1024^3$, $2048^3$ and $4096^3$ end at approximate times $150$, $300$, $600$ and $1100$ 
	respectively.
	The error bands mark the 	1-$\si$ and 2-$\si$ variations between realisations.
	Note that there is velocity data for only one simulation with $s=0$ and $N=4096$ in the matter era. The   simulation times are given in units where $\phi_0=1$.
	}
\label{fig:V2_lag}
\end{figure}

In Fig.~\ref{fig:v2_dx} we show the Lagrangian-weighted equation of state velocity estimator, averaged over the time intervals detailed in Tables 
\ref{t:simpars} and \ref{t:simpars_dx}, as a function of $\LatSpa/\Swid$, to investigate the importance of lattice spacing on this quantity. Note that only one of the 7 matter era $s=0$ runs had the mean square velocity recorded, and the error bar is estimated from the fluctuation in $\vAv^2$ over the time ranges in that run. This is likely to be an underestimate, as fluctuations within a run are highly correlated, and fluctuations between runs are much higher.

Table \ref{t:slopes_dx} shows the slopes of the fits, demonstrating that there is a significant dependence of the mean square velocity
on the lattice spacing.

There is also a less pronounced, but greater than 2$\si$ dependence on $\Swid/\xi$ in the radiation era, 
as can be seen in Table \ref{t:slopes_xi}.  
We perform an extrapolation with a joint fit to $\Swid/\xi$ and $\LatSpa/\Swid$, 
using the $s=1$ and the resolution test run data only.  
Including the $s=0$ data would bias the fit to towards data from $\LatSpa = 0.5$, where lattice effects are most important. 
The results are shown in Table \ref{t:final_fits}.

The value of $\vAv^2$ at $\LatSpa/\Swid = 0.5$ is approximately 20\% lower that the continuum extrapolation of $\vAv^2$, which quantifies the effect of coarser lattice spacings on the mean square string velocity.  An effect is to be expected, as strings on a lattice suffer a velocity-dependent retarding force \cite{Hindmarsh:2014rka}. We find that $\vAv^2$ decreases somewhat over the course $s=1$ simulations, as can be seen from the difference between the ``early'' and ``late'' values of $\vAv^2$ in Fig.~\ref{fig:v2_dx}. 
The decrease, due to the decreasing ratio $\LatSpa/\Swid$, limits the time span over which unequal time correlators can easily be taken: in \cite{Daverio:2015nva} we limited the time span to [600, 800] in the matter era and [450,600] in the radiation era,  during which time the ratio $\LatSpa/\Swid$ remains greater than 2. 

On the other hand, the $s=0$ mean square velocity remains constant throughout a simulation, as the ratio $\LatSpa/\Swid$ remains constant.  We can therefore take unequal time correlators over a much wider range of times, while recognising that corrections to the overall amplitude might need to be made so that they agree with the $s=1$ UETCs near equal times.  The details of how we do this are given in \cite{Daverio:2015nva}.

\begin{figure}[h]
\centering
\includegraphics[width=.4\textwidth]{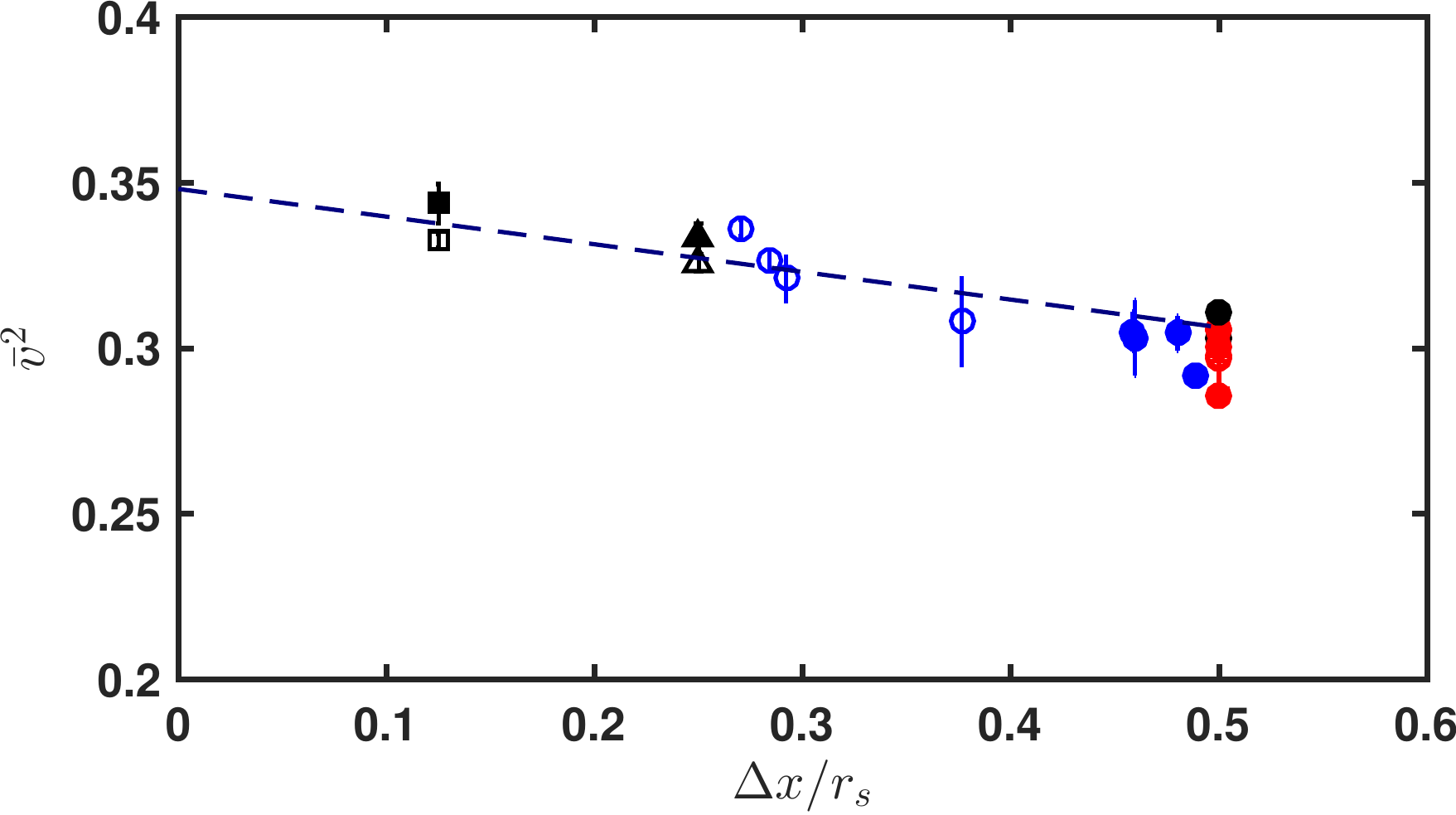}
\includegraphics[width=.4\textwidth]{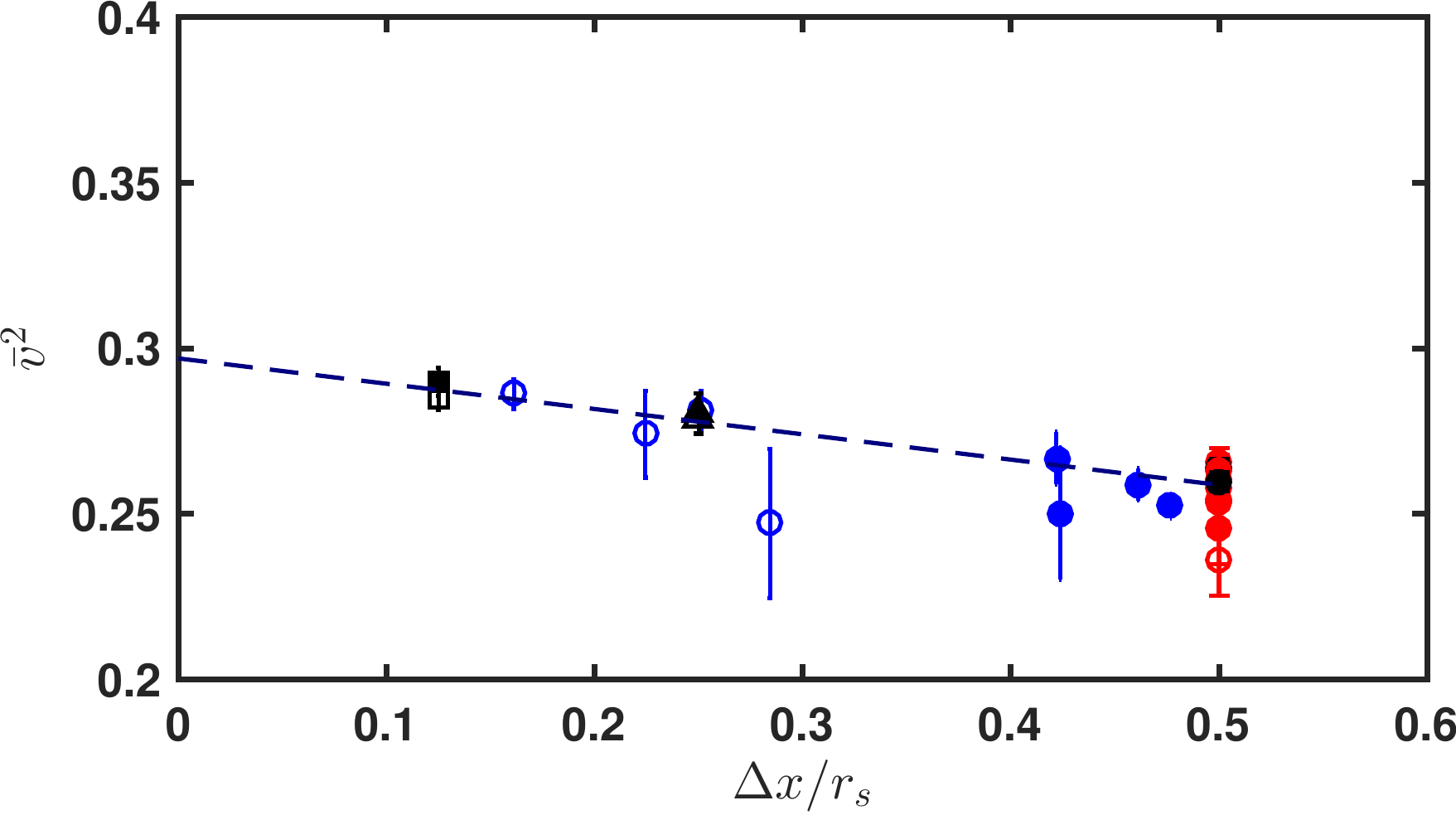}
\caption{Mean square string velocity from the equation of state estimator $\vAv^2_{w,\mcL}$ in radiation (top) and matter (bottom) eras, 
plotted against lattice spacing $\LatSpa$ in units of the string width $\Swid$.
The same colour and symbol code as Fig.~\ref{fig:xidot_lag} is used.
}
\label{fig:v2_dx}
\end{figure}

\begin{table}
\begin{tabular}{llcccc}
\hline
\hline
  Model & era & Dataset & $\dot\xi$         & $\vAv^2$          & $\tilde\ep$ \\
\hline
AH & Rad. & s1,R & $-0.02 \pm  0.06$ & $-0.08 \pm  0.01$ & $ 0.02 \pm  0.08$ \\
AH & Mat. & s1,R & $-0.01 \pm  0.03$ & $-0.08 \pm  0.01$ & $ 0.07 \pm  0.05$ \\
\hline
\hline
\end{tabular}
\caption{\label{t:slopes_dx}
Slopes of linear least squares fit of $\dot\xi$, $\vAv^2$ (computed from the Lagrangian-weighted equation of state) and $\Srep$ 
to the ratio $\LatSpa/\Swid$, where $\LatSpa$ is the lattice spacing 
and $\Swid$ the string width. 
Fits are performed on the  $s=1$ and resolution test run data.
Figs. (\ref{fig:v2_dx}) and (\ref{f:Srep_dx}) show the data for $\vAv^2$ and $\Srep$.
}
\end{table}

\begin{table}
\begin{tabular}{llcccc}
\hline
\hline
  Model & era & Dataset & $\dot\xi$         & $\vAv^2$          & $\tilde\ep$ \\
\hline
AH & Rad. &  s0l & $ 2.57 \pm  0.43$ & $ 1.59 \pm  0.62$ & $ 2.45 \pm  0.76$ \\
AH & Rad. &  s1l & $ 2.25 \pm  0.56$ & $ 1.04 \pm  0.52$ & $ 1.28 \pm  1.75$ \\
AH & Mat. &  s0l & $ 2.19 \pm  0.13$ & $ 0.53 \pm  0.48$ & $ 0.81 \pm  1.19$ \\
AH & Mat. &  s1l & $ 1.15 \pm  0.52$ & $ 0.82 \pm  0.56$ & $-2.27 \pm  1.92$ \\
\hline
\hline
\end{tabular}
\caption{\label{t:slopes_xi}
Slopes of linear least squares fit of $\dot\xi$, $\vAv^2$ (computed from the Lagrangian-weighted equation of state) and $\Srep$ 
to the ratio $\Swid/\xi$, where $\Swid$ is the string width and $\xi$ is the mean string separation. 
Fits are performed separately on the $s=0$ and $s=1$ data.
}
\end{table}

\section{Radiative efficiency parameter}

The fact that the strings scale means that there must be an energy loss mechanism with a special dependence on the comoving string separation $\xi$. 
Assuming that we have a method for picking out the string contribution to the energy-momentum, we can denote this contribution $\Sem_{\mu\nu}$.
The string separation parameter $\xi$ is then defined as 
\ben
\frac{\mu}{\xi^2} = - \frac{1}{\vol}\int d^3 x \Sem_{00},
\een
where $\vol$ is a volume factor. 
Covariant conservation of energy then implies that we can write
\ben
\frac{\mu}{\xi^2} \left( -2 \frac{\dot \xi}{\xi} + \frac{\dot{a}}{a} ( 1 + 3 \Seos) \right) = - \ep,
\een
where $\Seos$ is the string equation of state parameter $\Sem_{ii}/3\Sem_{00}$, and $-\ep$ is the energy density loss rate of the string network.
Hence
\ben
\label{e:SrepDef}
 \Srep \equiv \frac{\ep \xi^3}{\mu} = \left(2{\dot \xi} - \frac{\dot{a}}{a}{\xi} (1 + 3\Seos) \right)
\een
is a constant when the string network is scaling. We call the dimensionless parameter $\Srep$ the radiative efficiency parameter, which represents the fraction of the string network's energy radiated in a time interval $\xi$.  In order to achieve scaling, the radiative efficiency must be independent of the string separation parameter $\xi$.

\begin{table}
\begin{tabular}{llccc}
\hline
\hline
Model & era &  $\dot\xi$         & $\vAv^2$          & $\tilde\ep$ \\
\hline
AH & Rad. & $0.196 \pm 0.005$ & $0.370 \pm 0.011$ & $0.283 \pm 0.029$ \\
AH & Mat. & $0.226 \pm 0.006$ & $0.309 \pm 0.011$  & $0.185 \pm 0.043$  \\
\hline
NG & Rad. &   $0.15$ & $ 0.40 $ & $0.179$ \\
NG & Mat.  &   $0.17$ & $ 0.35 $ & $0.103$ \\
\hline
\hline
\end{tabular}
\caption{\label{t:final_fits}
Summary table of asymptotic values of $\dot\xi$, $\vAv^2$ (computed from the Lagrangian-weighted equation of state) and $ \Srep$.
The values of $\dot\xi$ are inferred from a linear least squares fit of the s1l dataset to the ratio $\Swid/\xi$,
extrapolated to infinite string separation. 
The values of $\vAv^2$ are inferred from a linear least squares fit of the s1 and R datasets 
to $\Swid/\xi$ and $\LatSpa/\Swid$, extrapolated to both zero lattice spacing and infinite string separation.
The values of $\Srep$ are inferred from a simple weighted mean of the s1 and R data. 
For comparison, we also show the equivalent values obtained from Nambu-Goto simulations \cite{BlancoPillado:2011dq}. 
}
\end{table}

In order to explore the radiative efficiency of Abelian Higgs strings, we define the string energy-momentum tensor as the Lagrangian weighted energy-momentum,
\ben
\Sem_{\mu\nu} = - T_{\mu\nu}\Ltilde,
\een
which is easily computed in a field simulation. 
Hence
\ben
\label{e:Srep}
\Srep  = 2\left({\dot \xi} - \frac{\dot a}{a}\xi \vAv^2_{w,\mcL}\right),
\een
where $\vAv^2_{w,\mcL}$ is the mean square string velocity computed from the Lagrangian-weighted equation of state (\ref{e:v2_w_lag}).

\begin{figure}[h]
\includegraphics[width=0.49\textwidth]{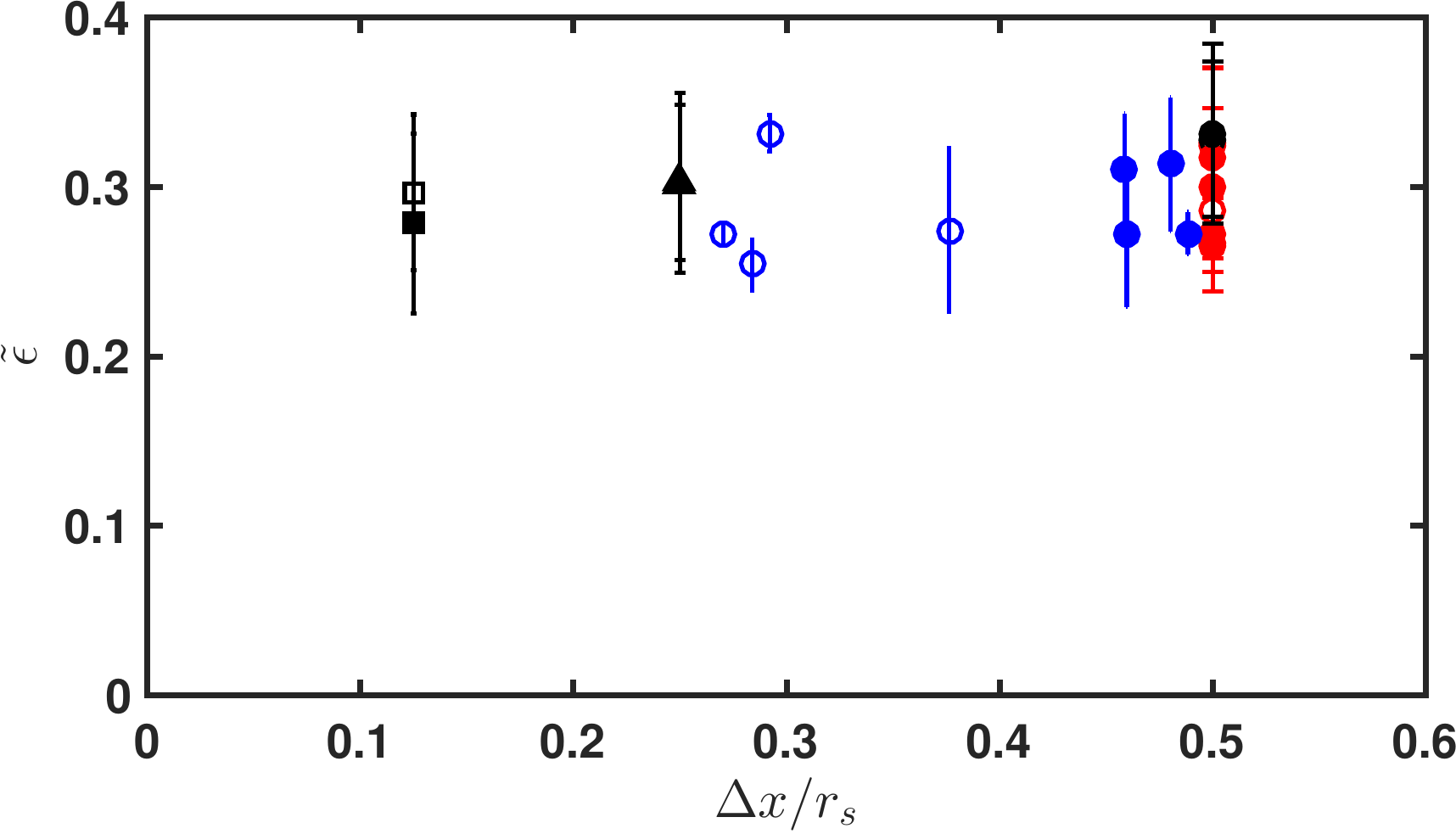}
\includegraphics[width=0.49\textwidth]{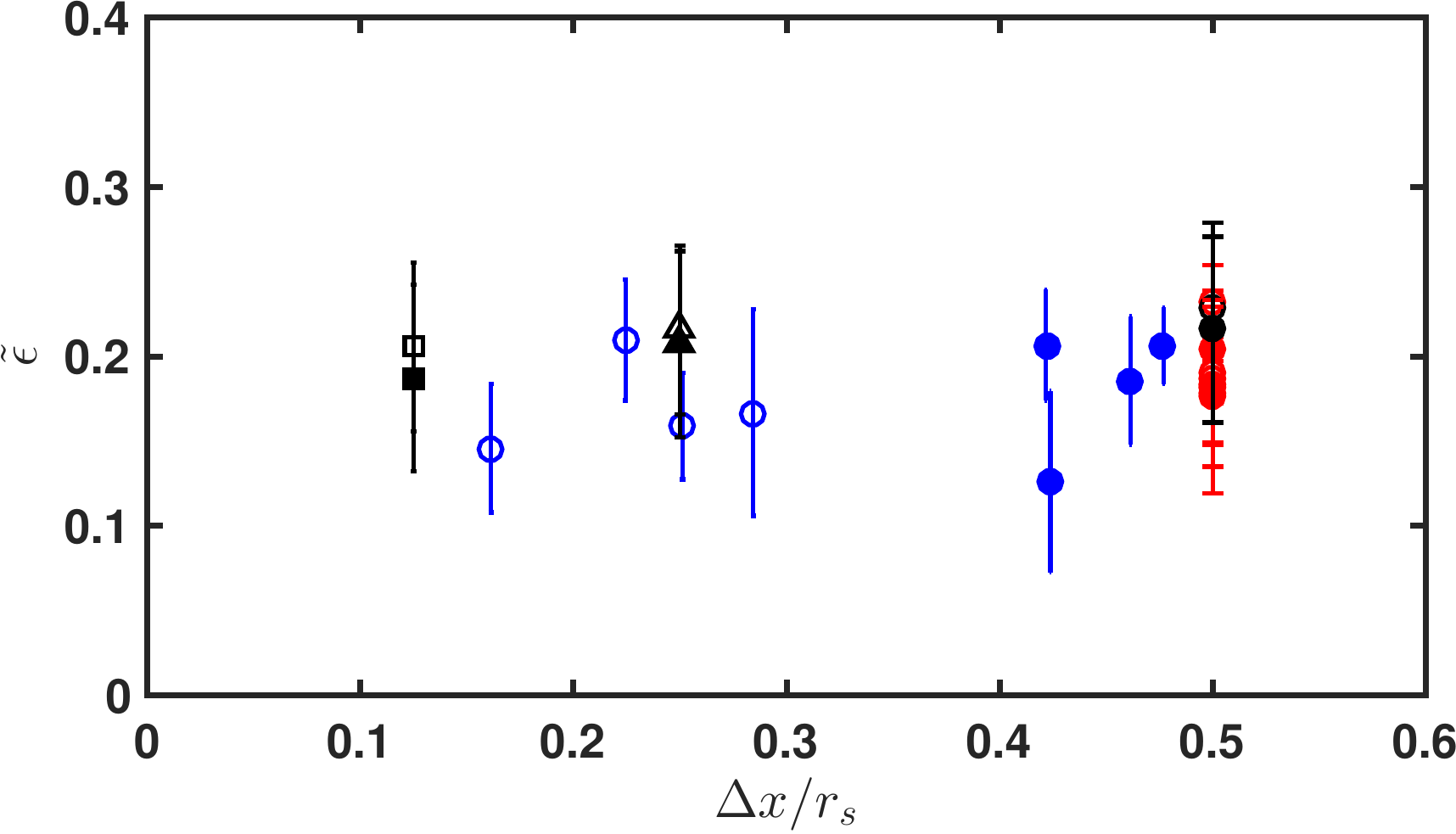}
\caption{\label{f:Srep_dx}Radiative efficiency parameter $\Srep$ (\ref{e:Srep}) against the ratio of lattice spacing to string width for radiation (top) and matter (bottom) eras. 
The same colour and symbol code as Fig.~\ref{fig:xidot_lag} is used.}
\label{fig:RadEffParDx}
\end{figure}

In Fig.\ \ref{f:Srep_dx} we plot the string radiative efficiency parameter, estimated from (\ref{e:Srep}), 
against the ratio of the lattice spacing to string width $\LatSpa/\Swid$,  for all simulations.  
The error bars are estimated from the uncertainties in $\dot\xi$ and $\vAv^2$, 
assuming that they are fluctuating independently.  
It is clear from Table \ref{t:slopes_dx}  
that there is little dependence on $\LatSpa/\Swid$.
Hence the radiative efficiency parameter is essentially independent of how well the string is 
resolved over the values of the lattice spacing chosen, which is strong evidence that the radiation is not a lattice artefact. 

In Fig.\ \ref{f:Srep_xi} we plot the string radiative efficiency parameter  
against the ratio of the string width to the string separation $\Swid/\xi$,  for all simulations.   
It is clear that there is also little dependence on $\Swid/\xi$: 
the values of $\Srep$ cluster around 0.2 in the matter era and 0.3 in the radiation era. 
We give the weighted means
in Table \ref{t:final_fits}

\begin{figure}[h]
\centering
\includegraphics[width=.49\textwidth]{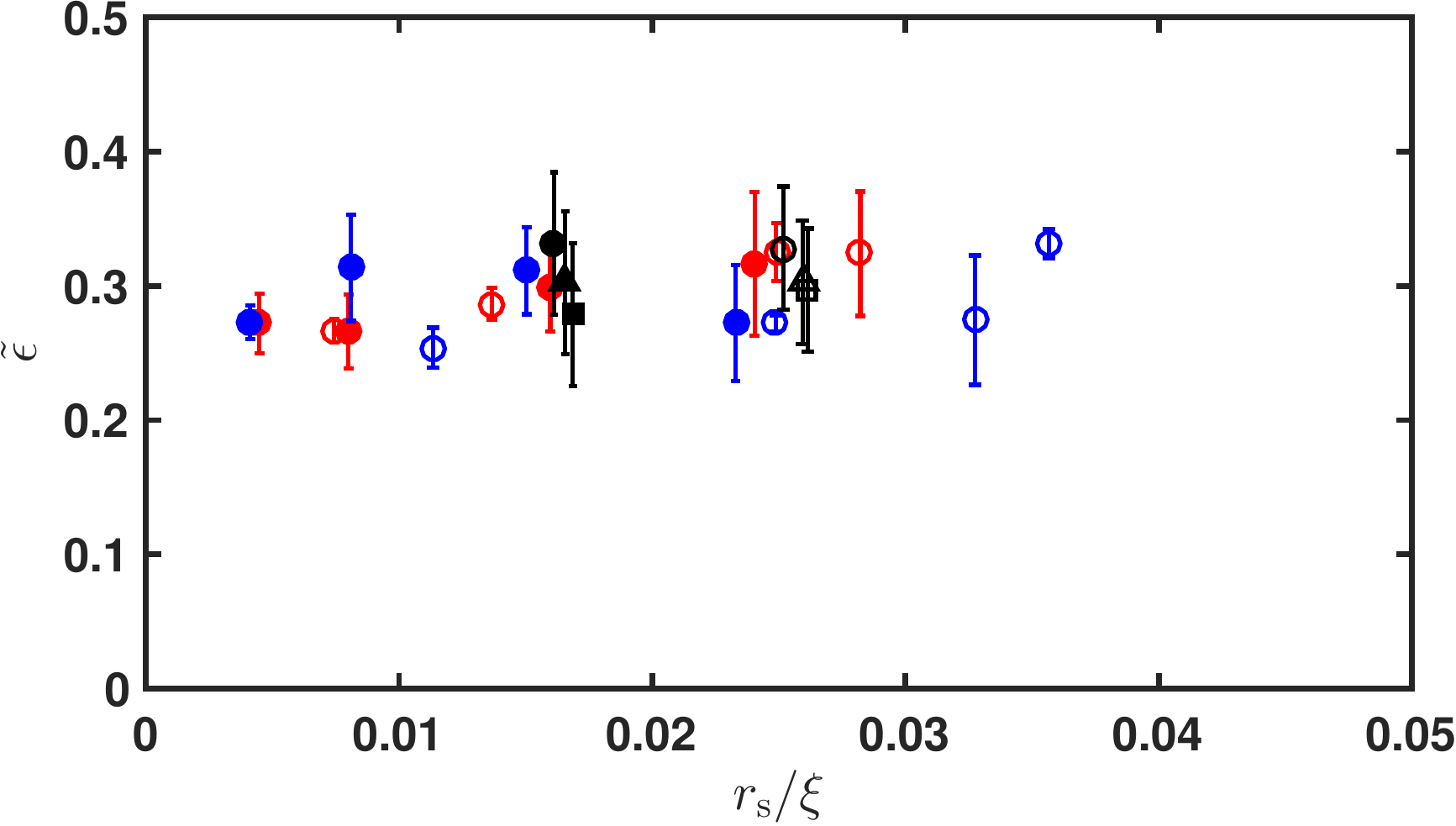}
\includegraphics[width=.49\textwidth]{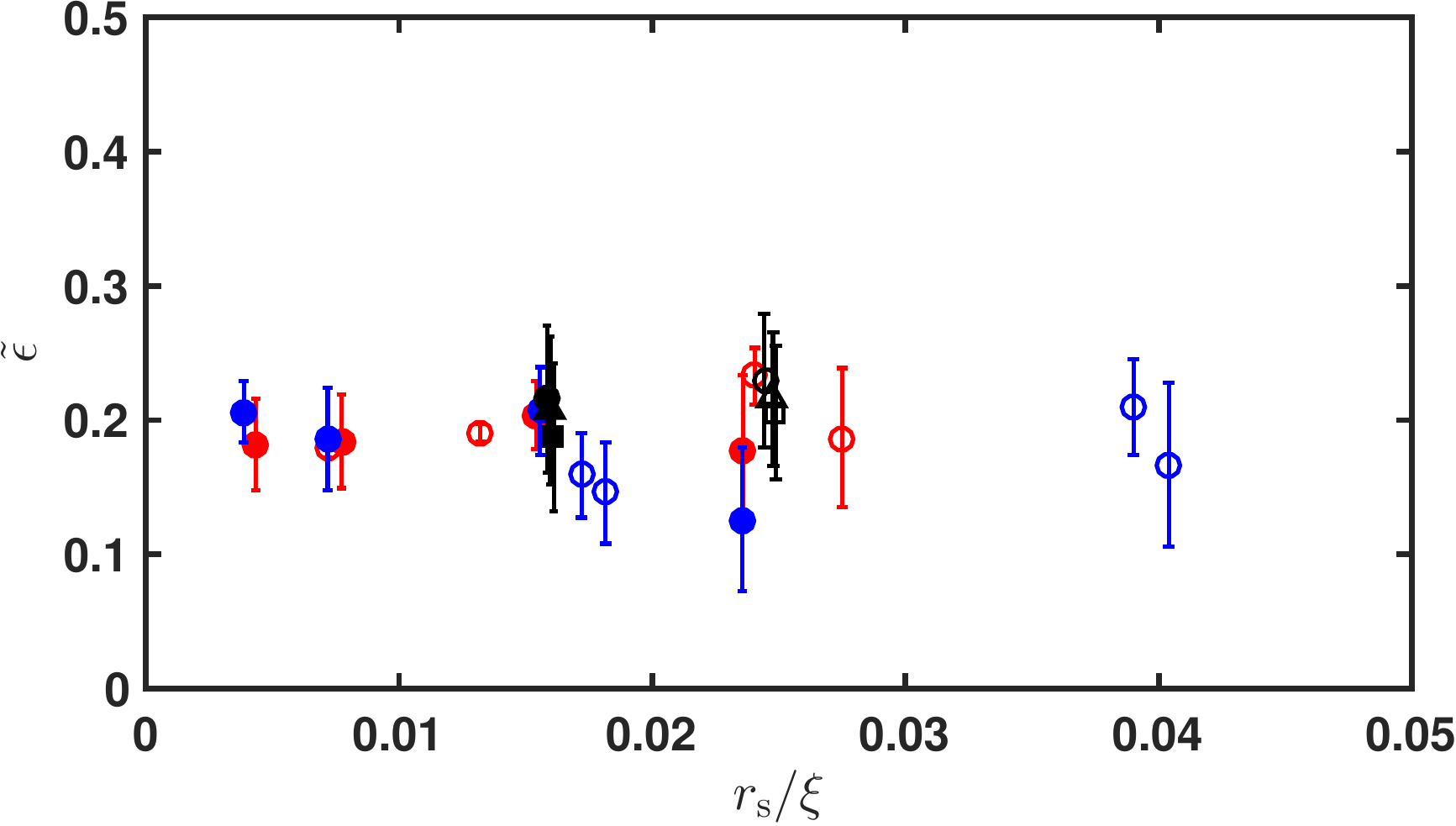}
\caption{\label{f:Srep_xi}Plots of the radiative efficiency parameter $\Srep$ (\ref{e:Srep})
	versus $\Swid/\xi$, in radiation (top) and matter (bottom) eras.
	The same colour and symbol code as Fig.~\ref{fig:xidot_lag} is used.
The mean values of $\Srep$ are listed in Table \ref{t:final_fits}.
 }
\label{fig:RadEffPar}
\end{figure}

It can be seen that the radiative efficiency parameter is lower in the matter era, where the mean square velocity is also lower.  This is consistent with the behaviour of the radiative efficiency in domain wall networks \cite{Martins:2016ois,Martins:2016lzc}.

It can also be seen that the simulations using the core growth method do not differ significantly in their radiative efficiency.

For comparison, we also give values obtained from Nambu-Goto simulations \cite{BlancoPillado:2011dq} 
of the equivalent observables, which are measured on string which is not in a stable (non-self-intersecting) loop.
The equivalent of the radiative efficiency parameter is calculated from the integral over the loop production function $\mathcal{P}$ 
(see Eqs.~9 to 11 in \cite{BlancoPillado:2011dq} ), and from $\ga = \dot\xi$, with $\Srep_\text{NG} = \ga^3\mathcal{P}$.  

\section{Discussion and Conclusion}

\subsection{Results summary}

In this paper we study in detail large-scale simulations of Abelian Higgs cosmic strings. 
We use simulations of standing waves to validate estimators of the principal global observables 
$\dot\xi$ (mean string separation in units of the horizon length) and $\vAv^2$ (mean square 
string velocity). 
We introduce  a new dimensionless parameter $\Srep$, the string radiative efficiency parameter, which quantifies 
the transfer of energy from strings to classical scalar and gauge radiation.  
We perform continuum limit and large-time extrapolations or network simulations to check the effects of lattice spacing and finite volume, 
and to provide estimates of the asymptotic values of the mean string separation, mean squared velocity, and radiative efficiency. 
We find that, contrary to naive expectation, the strings continue to radiate efficiently even when the string separation 
is much larger that the string width $\Swid$.

Standing wave solutions allow us to compare estimators of string length and string velocity, 
as the string follows a computable Nambu-Goto trajectory. 
Out of the string length estimators, the one based on the Lagrangian-weighted energy (\ref{e:ell_lag}) was closest to the standing wave prediction. 
For velocity estimation, the equation of state estimator $\vAv^2_{w,\mcL}$ given in Eq.\ (\ref{e:v2_w_lag}) performed best. 
Velocity estimators based on the positions of plaquettes with non-zero winding were consistently biased low. 

For network simulations, we expect to see scaling behaviour: that is, dimensionless global observables 
like $\dot\xi$ and $\vAv^2$ should tend to constants. This expectation is borne out: in all simulations, we find that the string separation $\xi$ is 
proportional to the conformal time (horizon length) to a good approximation after an initial period of relaxation, 
and that the mean square velocity is approximately constant, with values around $0.3$.
In detail however, there is variation between simulations, due to lattice spacing and finite size effects.

We examine the lattice spacing dependence of the observables $\dot\xi$, $\vAv^2$ and $\Srep$, 
as measured using the quantities described above. 
We performed simulations with a range of ratios of the lattice spacing $\LatSpa$ to the string width $\Swid$, including 
a set of three simulations with identical initial conditions 
resolved on grids with $\LatSpa/\Swid = 0.5$, $0.25$ and $0.125$.
The lattice spacing dependence is quantified in Table \ref{t:slopes_dx}.

We find that the mean square velocity shows significant lattice-spacing dependence, 
with a reduction of $\vAv^2$ of about 20\% relative to the continuum extrapolation at a lattice spacing $\LatSpa/\Swid = 0.5$.
There is no evidence of lattice spacing dependence in the measured values of $\dot{\xi}$, or 
in the radiative efficiency parameter $\Srep$. This indicates that the principal effect of the lattice is to slow the strings down, 
rather than to trigger radiation.
 
The relevant limit for the exploitation of our simulations in a cosmological context 
is the large time and large volume extrapolation, which is equivalent to $\Swid/\xi \to 0$.
We quantify the dependence of the global observables on $\Swid/\xi$ Table \ref{t:slopes_xi}, 
finding that only $\dot\xi$ varies significantly through the range explored by our simulations.

We give the values of $\dot\xi$, $\vAv^2$ and $\Srep$ in Table \ref{t:final_fits}, 
along with uncertainties. 
The mean string separation per horizon length $\dot\xi$ is derived from extrapolations of a least 
squares fit against $\Swid/\xi$, while the 
mean square velocity is derived from an extrapolation of a simultaneous fit to $\Swid/\xi$ and $\LatSpa/\Swid$.
The radiation efficiency parameter is a simple weighted average.
These are our estimates for the physical values for cosmic strings in the Abelian Higgs model.

\subsection{Implications for cosmological simulations}

The 4k simulations, with $N=4096$ points per side, form the basis of our recently published Cosmic Microwave Background 
power spectra for cosmic strings \cite{Lizarraga:2016onn}.
The important network observables from which the power spectra are calculated 
are the unequal time correlators (UETCs) of the energy-momentum tensor \cite{Daverio:2015nva}. 
They depend on the global quantities $\xi$ and $\vAv^2$.
In particular, the overall normalisation is proportional to $1/\xi$, which at large times can be written $1/\tau\dot\xi$.
The vector correlation is particularly sensitive to the mean square velocity, being proportional to $\vAv^2$.

In simulations with $s=1$ the string cores shrink in comoving coordinates, so that the effective resolution is higher 
than $\LatSpa/\Swid = 0.5$ for most of the simulation, and the effect of the grid on the string velocity is suppressed. 
For this reason we limited our determination of the near-equal time region of the UETCs 
to a range of times during which $\LatSpa/\Swid \lesssim 0.25$ 

For larger time ratios, we used $s=0$ simulations, scaled to match
the $s=1$ simulations. Our results in this work show that the different amplitudes of the UETCs
can be connected to the different scaling values of $\dot\xi$ and $\vAv^2$.  
However, in both cases, the extrapolated limit is compatible within error bars with the values obtained for the 4k simulations used in \cite{Daverio:2015nva}.
This means that the core growth method does not introduce significant systematic errors, 
although care must be taken to resolve the string core adequately in the comoving coordinates of the simulation.

Based on these results we conclude that the simulations used in \cite{Daverio:2015nva} show the scaling necessary to translate the measured UETCs to cosmological scales, and that their global observables $\dot\xi$ and $\vAv^2$ are sufficiently close to their asymptotic values  
to correctly represent the large-time Abelian-Higgs string dynamics.

\subsection{Nambu-Goto approximation}

It is common to assume that cosmic strings can be described as infinitely thin relativistic strings, or Nambu-Goto strings, as $\Swid/\xi \to 0$.
Our simulations can be used to test this assumption, 
and to compare observables against the most recent simulations of networks of Nambu-Goto (NG) strings 
\cite{BlancoPillado:2011dq,Blanco-Pillado:2013qja}.
The NG large-separation limits for $\dot\xi$ and $\vAv^2$ are quoted in Table  \ref{t:final_fits} 
for comparison with the results for Abelian Higgs strings, although it should be noted that the 
NG data is calculated for ``long'' (length greater than $\xi$) strings only.  

As discussed in Section \ref{s:AHprops}, energy conservation for NG strings is equivalent to length conservation, and 
so the total length of string in an NG network remains constant. Scaling manifests itself as the conversion of 
long strings to loops with lengths $\ell \lesssim \xi$. Most of the loops have sizes around the initial 
correlation length, although the energy (i.e.~string length) is distributed over a wide range of scales, 
with an approximately constant fraction of energy being transferred to stable loops of size $\ell \sim \xi$.
There is non-trivial small-scale structure on the string network \cite{Polchinski:2006ee,Polchinski:2007rg}, 
as measured by the tangent vector correlator $\vev{\bXp(\si) \cdot \bXp(0)}$ at short distances.

The fact that the string length decreases in our simulations means that the NG approximation is failing.
However, if one excludes non-self-intersecting loops, the remaining string in a NG simulation behaves similarly to an Abelian Higgs network:  
the string separation in units of the horizon length, $\dot\xi$, and the mean square velocity $\vAv^2$ both tend to constants.
We see from Table \ref{t:final_fits} that the mean string separation is lower and 
that the mean square velocities for NG simulations are higher, by about 15\%.

The reason for these differences may lie in the NG approximation's neglect of classical radiation, which 
one can see is about 70\%  more efficient than loop production (last column of Table \ref{t:final_fits}). 
One would therefore expect that the field theory strings would have a lower density and so a larger separation. 
One might also expect that the radiation reaction would slow the field theory strings down. 

The main difference between the NG simulations of 
\cite{Martins:2005es,Ringeval:2005kr,BlancoPillado:2011dq,Blanco-Pillado:2013qja,Blanco-Pillado:2015ana}
and those of strings in the Abelian Higgs model
is the population of stable non-self-intersecting loops.
As the massive radiation channel is excluded by the NG approximation, 
loops have no way of decaying. 
In the field theory, a loop of string of size $\ell$ 
radiates scalar and gauge radiation, 
and disappears in a time of order $\ell$ \cite{Hindmarsh:2008dw}.
The global efficiency of this process is described by the radiative efficiency parameter $\Srep$. 
It would be very interesting to determine where the radiation is coming from: is the radiative efficiency 
the same for long strings as for loops, or do long strings primarily lose energy into loops
which then are responsible for most of the radiation. 
We will examine how the radiative efficiency depends on loop size in a future publication.

It was shown in \cite{Moore:2001px} that a substantial proportion of the long string 
energy goes into loops, and it was speculated that these loops would eventually 
be stable in a large enough simulation. We see no evidence for a population of 
stable loops in field theory strings: they would show up as a departure from 
linear behaviour in the mean string separation. 
We have recorded the position data for all strings, and we will check every loop 
looking for stability in the future publication mentioned above.

\subsection{Radiation from strings}

Field theory strings maintain their scaling configuration by emitting radiation in the form of massive fields.
Naively one would expect that the radiative loss decreases exponentially as the string separation increases, 
as the string separation should be related to the mean curvature radius of the strings. 
Indeed, this is the case for standing wave configurations carefully prepared by cooling the 
field configuration around a sinusoidal string position  \cite{Olum:1999sg}.  

The initial fields for our string networks are cooled in an identical way to those for the standing waves,  
and so in the initial phase of evolution the fields are as smooth as those which comprise the standing wave.  
Yet, after a short interval, the strings in a network start to lose energy, unlike standing waves.

We quantify the energy loss from a string network with the radiative efficiency parameter $\Srep$, 
defined as the fraction of the energy in a string network radiated in a time interval $\xi$. 
It is conceptually similar to the loop chopping efficiency $\tilde{c}$ introduced in modelling of Nambu-Goto networks \cite{Martins:2000cs}, 
which quantifies the transfer of energy from long strings into loops.
In the context of domain wall simulations it has been also called 
the energy loss function \cite{Martins:2016ois,Martins:2016lzc}.

We have conclusively demonstrated that the radiation is not a lattice artefact. 
We also find no significant dependence to $\Srep$ on $\Swid/\xi$, 
in the range explored by our simulations of $0.05 \gtrsim \Swid/\xi \gtrsim 0.003 $. 
In other words, the power per unit length of string decreases as $1/\xi$ rather than exponentially. 
It is this behaviour, encoded in the constancy of $\Srep$, 
that allows Abelian-Higgs string networks to scale.

It is rather remarkable that the string network is 
able to transport energy from the scale $\xi$ into massive radiation with wavelength 
$\Swid$, a factor of over 300 at the end of the largest simulations.  
The fact that the energy loss mechanism is essentially independent of the ratio $\Swid/\xi$ means that it 
cannot be perturbative radiation, 
which would decrease exponentially with $\xi/\Swid$ \cite{Olum:1999sg}.  
Hence, strings in a dynamic field theory network violate one of the assumptions behind the 
Nambu-Goto approximation. 
We believe that the incorrect assumption is that the field configuration in the local rest frame of the string is 
precisely that of the Nielsen-Olsen vortex with curvature $\xi^{-1}$.  
As discussed in \cite{Hindmarsh:2008dw}, small-scale structure exists down to a length of order the 
string width $\Swid$, and perturbs the field configuration
in modes of frequency up the mass scale of the scalar and gauge fields, which can 
leave the string as radiation.
The transport of energy over a large range of scales resembles turbulence, but a detailed understanding is still lacking.
A visualisation of isosurfaces of energy density in a $1024^3$ radiation era $s=0$ run, 
illustrating the complexity of the field configuration, 
can be found at \cite{Visu}.

\subsection{Conclusions}

To summarise, we have exhaustively investigated lattice spacing and finite size effects in the simulation of cosmic string 
networks in the classical Abelian Higgs model.
We conclude that our $N=4096$ (``4k'') simulations exhibit scaling over a wide enough range of scales 
to allow extrapolation to cosmological scales at late times, and that lattice spacing effects are well under control.
We have quantified the uncertainties in the extrapolated global quantities $\dot\xi$, the mean string separation as a fraction of the 
horizon, and $\vAv^2$, the mean square string velocity.

The observed scaling of the Abelian Higgs string network implies the existence of an energy loss mechanism that 
transports energy from large scales to small,
and we show that the efficiency of this energy loss mechanism,
defined as the fractional energy loss in the time $\xi$, 
is independent of the ratio of string separation $\xi$ to string width, implying 
that it can be extrapolated to large $\xi$ and hence large times.

We conclude that our simulations are good evidence that  
scaling due to energy loss by classical radiation of massive fields 
is the correct scenario for field theory strings. 
The mechanism by which this radiation is produced is still not well understood, but the 
same mechanism is presumably at work producing scaling in domain wall networks in two 
and three dimensions \cite{Press:1989yh,Garagounis:2002kt,Borsanyi:2007wm,Leite:2011sc,Martins:2016ois}.

Furthermore, the ability of strings to transport energy from large scales to small is clear even in Nambu-Goto strings, 
where most loops are produced with size of order the initial correlation length \cite{Ringeval:2005kr,BlancoPillado:2011dq}, 
maintaining small-scale structure on the string network \cite{Martins:2005es}. 
We will investigate small-scale structure and loops in Abelian Higgs string networks in a future publication.

Our results showing the linear growth of the string separation imply that field-theory strings do not form large stable loops. 
Hence modelling based on the Nambu-Goto approximation exaggerates the density of string loops, 
and therefore the amount of gravitational radiation emitted. 
This means that gravitational wave bounds based on 
radiation from stable loops \cite{Sanidas:2012ee,Blanco-Pillado:2013qja,Aasi:2013vna} 
are likewise overestimated. 
The gravitational wave flux from Abelian Higgs strings is yet to be calculated, but 
one would expect the resulting spectrum to be more like that from global cosmic strings \cite{Figueroa:2012kw}.

The implication is that there are no significant gravitational wave constraints on cosmic strings in field theories.
Stronger constraints can be derived from bounds on high-energy particle emission \cite{Mota:2014uka}, 
or Cosmic Microwave Background fluctuations \cite{Lizarraga:2016onn} if the fields making the 
string are decoupled from the Standard Model.

On the other hand, the Nambu-Goto approximation can be used to describe the large-scale properties of string networks, 
as are relevant for CMB perturbations \cite{Lazanu:2014eya,Lazanu:2014xxa}, with 20 -- 30\% accuracy.
However, any observables which depend on a population of long-lived loops 
cannot be reliably be computed for field theory strings in the Nambu-Goto approximation.

The Nambu-Goto scenario may eventually be recovered if 
there is a very mild decrease of the radiative efficiency on $\Swid/\xi$ 
undetected over the range explored in our simulations, which becomes 
significant over cosmological scales.

Our work does not directly impact on the use of the classical Nambu-Goto action to describe 
the dynamics of fundamental cosmic strings \cite{Sarangi:2002yt,Copeland:2003bj,Dvali:2003zj}, 
although our demonstration of the ability of strings to transport energy to the shortest length scale in the system 
and the solitonic aspects of D-strings \cite{Dvali:2002fi} may give cause for a re-evaluation of the Nambu-Goto approximation in that case too.


\begin{acknowledgments}
This work has been supported by two grants from the Swiss National Supercomputing Centre (CSCS) under project IDs s319 and s546. In addition  this work has been possible thanks to the computing infrastructure of the i2Basque academic network, the COSMOS Consortium supercomputer (within the DiRAC Facility jointly funded by STFC and the Large Facilities Capital Fund of BIS), and the Andromeda/Baobab cluster of the University of Geneva. 
We also acknowledge the use of NERSC computational facilities during their test phase, and thank Julian Borrill for help.
MH (ORCID ID 0000-0002-9307-437X) acknowledges support from the Science and Technology Facilities Council 
(grant number ST/L000504/1). DD and MK acknowledge financial support from the Swiss NSF. 
JL and JU acknowledge support from Eusko Jaurlaritza (IT-979-16) and the Spanish Ministry MINECO (FPA2015-64041-C2-1P). \end{acknowledgments}

\bibliography{CosmicStrings}

\end{document}